\documentclass[rmp,aps,nofootinbib,twocolumn]{revtex4}

\usepackage{graphicx}
\usepackage{hyperref}

\newcommand{\eps} {\varepsilon}
\newcommand{\nn} {\nonumber}
\newcommand{\vphi} {\varphi}

\begin{document}
\title{Colloquium: Phase transitions in polymers and liquids in electric fields}

\author{Yoav Tsori}
\email{tsori@bgu.ac.il}
\affiliation{Department of Chemical Engineering, Ben-Gurion University of the Negev,
84105 Beer-Sheva, Israel}


\begin{abstract}  

The structure and thermodynamic state of a system changes under the
influence of external electric fields. Neutral systems are characterized by
their dielectric constant $\eps$, while charged ones also by their charge
distribution.
In this Colloquium several phenomena occurring in soft-matter systems
in spatially uniform and nonuniform fields are surveyed and the role of the 
conductivity $\sigma$ and the 
linear or nonlinear dependency of $\eps$ on composition are identified.
Uniform electric fields are
responsible for elongation of droplets, for
destabilization of interfaces between two liquids, and for mixing effects in liquid mixtures.
Electric fields, when acting on phases with mesoscopic order,
also give rise to block copolymer orientation, to destabilization of 
polymer-polymer interfaces, and to order-order phase transitions.
The role of linear and nonlinear dependences of $\eps$ on composition 
will be elucidated in these systems. 
In addition to the dielectric anisotropy, existence of a finite conductivity leads to
appearance of
large stresses when these systems are subject to external fields and usually to
a reduction in the voltages required for the instabilities or phase transitions
to occur.
Finally, phase transitions which occur in nonuniform fields are described and
emphasis on the importance of $\eps$ and $\sigma$ is given.

\end{abstract}

\maketitle
\tableofcontents

\section{Introduction} \label{intro}

There are numerous ways the structure and phase of a system can be influenced by
external fields. The gravitational field can lead to
changes in structure and composition, and in many cases even to
phase transitions. However, while omnipresent on Earth, its effect
is usually quite small. A magnetic field has several
advantages,
but it is also weak unless the system is magnetic. Shear forces are effective in
changing the phase and structure of soft materials, but their presence is
undesired in many cases. Electric fields, on the other hand, have several
interesting properties: they have a strong effect even on neutral materials, they
can be switched on or off, and they are ideally suited for the nanoworld,
because an electric field increases linearly with a decrease in the electrode
size, for a given electrode potential.

The subject of this Colloquium is the influence that electric fields have on
interfaces between two liquids or polymers.
This is a broad subject and we do not aim at fully covering 
it but rather mention in this review the main effects.
An electric field deforms the interface and may give rise to a dynamical
instability. Similarly,
fields tend to orient ordered phases of heterogeneous polymeric materials 
in such a way as to minimize the
electrostatic energy. Uniform electric fields can also lead
to the {\it creation} of interfaces between liquids. 
Lastly, spatially nonuniform fields have a strong influence on the
thermodynamic properties of liquid and surfactant mixtures.

We adopt a continuum coarse-grained approach, where all quantities vary
smoothly enough to be described by continuous fields. 
In addition, interfaces between two pure immiscible liquids
are taken to be infinitesimally thin.
We seek to find the spatial distribution of liquid or polymer composition and
electric field.
The main ingredients are
the dielectric constant difference $\Delta\eps$, $\eps''$, and the conductivity
$\sigma$, to be defined below. The system's behavior 
depends on the these quantities and also on whether the fields are
spatially uniform or not. This approach neglects the molecular details and
therefore lacks accuracy, but it is general enough and captures the physical
mechanisms at play. 

Let us first recall few basic laws of electrostatics. The
electric field ${\bf E}$ is derivable from an electrostatic potential $\psi$,
such that
${\bf E}=-\nabla\psi$. The potential satisfies Poisson's equation
$\nabla\cdot(\eps\nabla\psi)=-\rho$, where $\rho$ is the free charge density and
$\eps$ is the medium dielectric constant. We assume throughout this article that
the system is isotropic, and therefore $\eps$ is not a tensor but a scalar.
The displacement field ${\bf D}$ is usually given by the linear
relation ${\bf D}=\eps{\bf E}$ and is the quantity conjugated to the electric
field. 

The component of ${\bf D}$ perpendicular to a sharp
interface between two materials with dielectric constants $\eps_1$
and $\eps_2$ is
continuous across the interface. Thus, the perpendicular component of ${\bf E}$
is discontinuous and is given by $\eps_1E_1=\eps_2E_2$,
where $E_1$ and $E_2$
are the electric fields in the perpendicular direction. In contrast, the
parallel component of ${\bf E}$ is continuous across the interface, while that
of ${\bf D}$ is not.
In systems where the charge on the conductors bounding the dielectric material
is given, ${\bf D}$ is the natural variable, and ${\bf E}={\bf E}({\bf D})$. The
electrostatic energy density of the pure dielectric is given by $F_{\rm
es}=(1/2)\int ({\bf D}^2/\eps){\rm d}^3r$.
Conversely,
in situations where the potential is specified on the set of bounding
conductors,  ${\bf E}$ is the natural variable,
and 
${\bf D}={\bf D}({\bf E})$. A Legendre transform then yields the electrostatic
energy,
\begin{equation}\label{f_es}
F_{\rm es}=-\frac12\int \eps{\bf E}^2{\rm d}^3r~.
\end{equation} 
Note the minus sign in front of the integral. Here is a simple argument for it; 
a full explanation can be found in \textcite{LL_electrodynamics}.
Our system is subjected to a fixed voltage. We can therefore imagine a large 
condenser of
capacitance $C$ and charge $Q$ connected to our system in parallel. 
The voltage imposed by this condenser is $V_0=Q/C$. Initially,
our small system, having a small capacitance $c\ll C$, is not charged (${\bf E}=0$
everywhere), and the total electrostatic energy is $U_1=(1/2) Q^2/C$. We now
connect the
two 
condensers, and charge $q\ll Q$ enters our system. We find $q=Qc/(c+C)$ 
and $U_2=(1/2) Q^2/(c+C)\simeq (1/2) Q^2/C(1-c/C)$. To first order in $c/C$,
the
change
in electrostatic energy is found to be $U_2-U_1=-(1/2) cV_0^2$. 
This argument shows that for a system under fixed voltage, the minus sign in 
Eq. (\ref{f_es}) correctly accounts for the work done by the external power supply.

In this article we are interested in liquids, liquid mixtures, and
block copolymers in electric fields. 
It is convenient to define an order parameter $\phi({\bf r})$,  a
spatially dependent dimensionless quantity denoting the relative composition of
one liquid or copolymer component, $0<\phi<1$. We denote by $\vphi$ the
variation in $\phi$ from the average value $\phi_0$,
\begin{eqnarray}
\phi({\bf r})&=&\phi_0+\vphi({\bf r})~,\nn\\
\langle \vphi({\bf r})\rangle&=&0~.
\end{eqnarray}  
The variation $\vphi$ induces a variation in the dielectric constant $\eps$. If
$\vphi$ is small enough, one may write a constitutive relation $\eps(\phi)$
as a Taylor series expansion to second order in $\vphi$,
\begin{eqnarray}\label{const_relation}
\eps(\phi)=\bar{\eps}+\Delta\eps\vphi+\frac12\eps''\vphi^2~.
\end{eqnarray}
$\bar{\eps}=\eps(\phi_0)$ and is the average dielectric constant if $\eps''$ is absent
from the expansion.
At the moment we consider ``neat'' dielectrics, which contain no dissociated
ions;
presence of salt will be allowed later. 
The electric field originates from the presence of a given arbitrary collection
of
conducting bodies at fixed, prescribed, potentials and/or charges. We define ${\bf E}_0$
as the electric field which is present in the system when $\eps$ is constant everywhere, 
$\eps=\bar{\eps}$. Variations in composition $\vphi$ lead to variations in $\eps$, and
since $\eps$ and ${\bf E}$ are coupled via Laplace's equation, one has variations in
electric field. 

We may thus write to quadratic order in $\vphi$
\begin{eqnarray}\label{E_expansion}
{\bf E}={\bf E}_0+{\bf E}_1\vphi+\frac12{\bf E}_2\vphi^2~.
\end{eqnarray} 
When we later refer to ``uniform electric fields'', we mean that ${\bf E}_0$ is
constant everywhere. This means that ${\bf E}_0$ actually originates from a
parallel-plate condenser or from nonideal planar electrodes lying far enough
from the point of interest, such that field inhomogeneities can be safely
neglected. Of course, even if the zeroth-order field ${\bf E}_0$ is uniform,
composition variations lead to field nonuniformities, as is evident in the
above expansion.

Let us look at the different terms in an expansion of the electrostatic energy density
[Eq. (\ref{f_es})] in powers of $\vphi$,
\begin{eqnarray}\label{f_es_expansion}
f_{\rm es}&=&-\frac12\bar{\eps}{\bf E}_0^2-\left(\bar{\eps}{\bf E}_0\cdot{\bf
E}_1+\frac12\Delta\eps{\bf E}_0^2\right)\vphi-\frac12(\bar{\eps}{\bf
E}_1^2\nn\\
&+&2\Delta\eps{\bf E}_0\cdot{\bf
E}_1+\frac12\eps''{\bf E}_0^2+\bar{\eps}{\bf E}_2\cdot{\bf
E}_0)\vphi^2+O(\vphi^3).~~~~
\end{eqnarray} 
(Note that energy densities are marked by lowercase letters.)

The first term on the right is an unimportant constant. The two terms in linear order of
$\vphi$ are inconsequential for the thermodynamic state of the system as long as the external
field ${\bf E}_0$ is uniform. To see this, one may write ${\bf E}_1$ as a sum of two
components: ${\bf E}_1=\alpha_\parallel{\bf E}_0+\alpha_\perp{\bf E}_{1\perp}$,
where ${\bf E}_{1\perp}$ is the component perpendicular to ${\bf E}_0$. If ${\bf E}_0$ is
uniform, one finds that $\alpha_\parallel$ is independent of ${\bf r}$ and 
$\int \vphi{\bf E}_1\cdot{\bf E}_0{\rm d}^3r=0$.
As a results, the linear term in $\vphi$ in Eq.
(\ref{f_es_expansion}) vanishes upon spatial integration, recalling that
$\langle\vphi({\bf r})\rangle=0$. When ${\bf E}_0$ varies in space, this is
no longer true since 
a dielectrophoretic force acts on the system. Several drastic
thermodynamic changes become possible, as is discussed in Sec.
\ref{sect_efips}.

The first and second terms in the second line of Eq. (\ref{f_es_expansion})
($\propto\vphi^2$) are important in uniform fields. The second term is
twice as large as the first one and opposite in sign, and the two sum to give a
free energy
contribution proportional to the dielectric contrast squared $+(\Delta\eps)^2$. This is a
free energy penalty for dielectric interfaces perpendicular to the external
field. We
explain how these terms give rise to a normal-field instability in liquids
(Sec. \ref{interf_instab}), to various orientation effects occurring in
mesoscopically ordered polymer 
phases (Sec. \ref{sect_bcp}), and to the phase behavior of
liquid
mixtures and block copolymer melts (Sec. \ref{sect_crit_eff}).
The so-called Landau mechanism term, proportional to $\eps''$, is
responsible for a modification of the liquid-vapor and liquid-liquid coexistences
in electric fields (see Sec. \ref{sect_crit_eff}).

\section{Normal Field Instability}
\label{interf_instab}

In this section we describe the interfacial instability occurring when an initially flat
interface separating two immiscible liquids or a liquid and a gas is subjected to a
perpendicular electric field.

\subsection{Dielectric interfaces}\label{droplet_elongation}

When an initially spherical liquid droplet is put in a uniform external electric field, it
elongates in the direction of the field. The degree of elongation is given as a balance
between electrostatic energy, preferring a long, needlelike, drop, and surface tension,
preferring a spherical object \cite{taylor1964}. 
\textcite{okonski1953} and later 
\textcite{mason1962} obtained the following expression for small deformations of drops:
\begin{eqnarray}\label{drop_elongation}
\frac{R_\parallel-R_\perp}{R_\parallel+R_\perp}=\frac{9}{16}\frac{R\eps_2 E_0^2}{\gamma}
\frac{(\eps_1-\eps_2)^2}{(\eps_1+2\eps_2)^2}~.
\end{eqnarray}
$\eps_1$ and $\eps_2$ are the dielectric constants of the drop and the embedding medium, respectively.
$R_\parallel$ and $R_\perp$ are the radii of the ellipsoid in the directions
parallel and perpendicular to the electric field of strength $E_0$ far from the
drop, $R$ is the unperturbed drop radius, and $\gamma$ is the interfacial tension
between the two liquids.
Clearly, the drop elongates in the field's direction. The deformation vanishes
in the absence of dielectric contrast, that is when $\eps_1=\eps_2$, and does
not depend on the
sign of
$E_0$. The case of a conducting drop is also given in the limit $\eps_1/\eps_2\to\infty$.

Some of the results of Allan and Mason for dielectric drops
showed an opposite behavior that could not be explained by the theory of neat
dielectric liquids -- in a few cases the drops became oblate rather than prolate.
These results led G. I. Taylor to propose his ``leaky dielectric'' model. Taylor
realized that even a small conductivity of the embedding liquid could lead to
significant changes to the drop shape. Indeed, the tangential stress cause by
the ionic flow leads to flow inside the drop \cite{taylor1966}. The solution of
the full electrohydrodynamic problem led him to suggest a discriminating
function $\Phi$ obeying
\begin{eqnarray}
\Phi=R(D^2+1)-2+3(RD-1)\frac{2M+3}{5M+5}~.
\end{eqnarray}
Here, $D$, $R$, and $M$ are the ratios of dielectric constant, resistivity, and viscosity 
of the outer liquid to those of the drop \cite{saville1997}. 
Drops are prolate when $\Phi>0$, oblate when $\Phi<0$, and spherical if $\Phi
=0$.
Qualitative agreement has been found between experiments and theory derived
from Taylor's initial study \cite{saville1997}.

\begin{figure}[h!]
\includegraphics[scale=0.45,bb=30 615 545 755,clip]{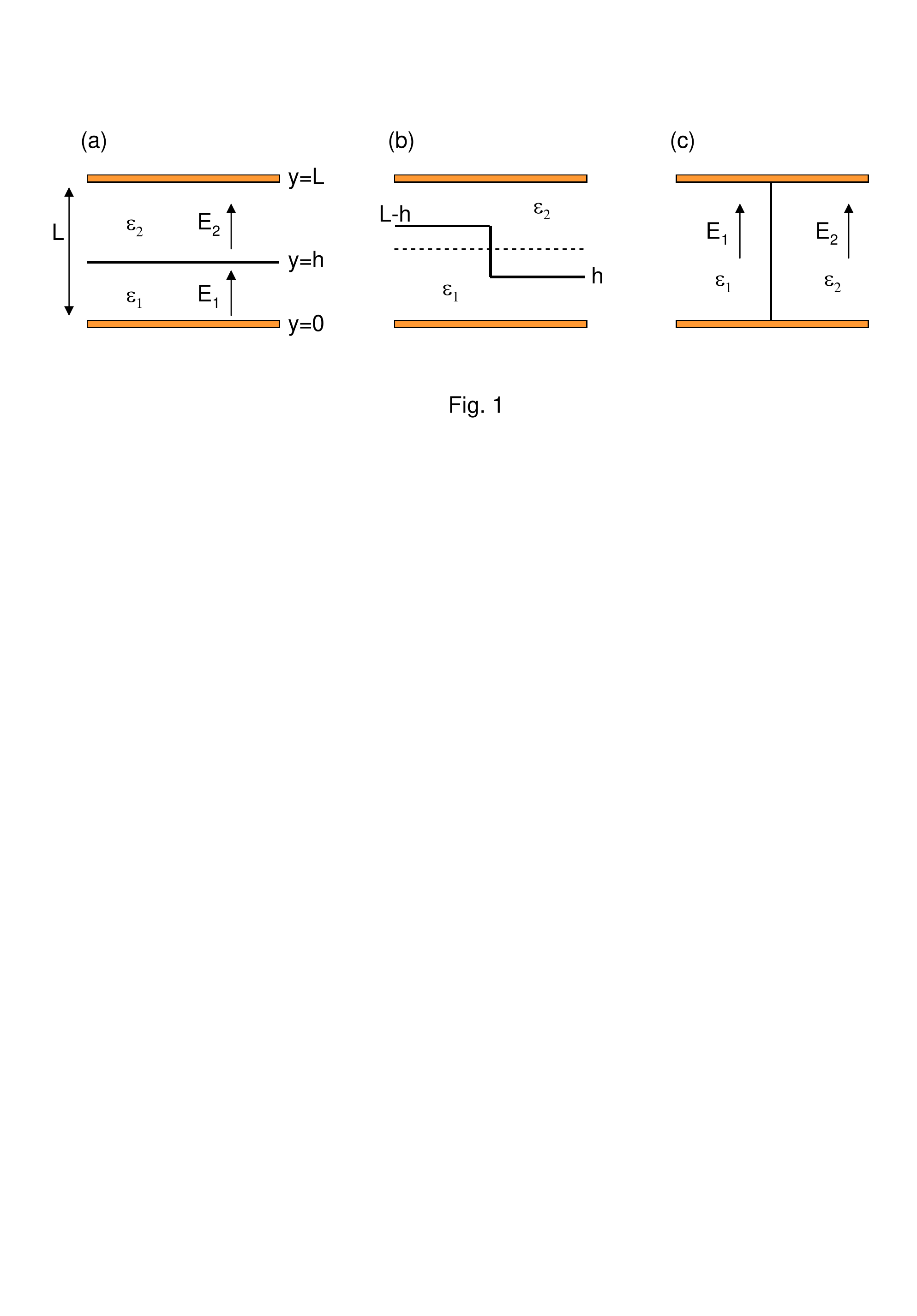}
\caption{Two liquids in uniform electric field. 
(a) Schematic illustration of two liquids with permittivities $\eps_1$
and
$\eps_2$ confined between two plates at separation $L$ and voltage difference $V$. The
pressure on the interface located at $y=h$ is given by Eq. (\ref{p_es}). (b) An initially
flat interface at $h=L/2$ can break into two parts. The electrostatic energy, given by Eq.
(\ref{es_energy_broken}), prefers $h=0$ or $h=L$. The preferred state is shown in (c).
}
\label{fig_liquid_para_perp}
\end{figure}

The role of residual conductivity, as understood by Taylor, will be further
highlighted in this review. Let us, however, return to pure dielectric fluids.
Consider this time two incompressible fluids denoted 1 and 2, confined in a
condenser with plate separation $L$, area $S$, and potential difference $V$ [see
Fig.~\ref{fig_liquid_para_perp}(a)]. The location of the flat interface 
between the fluids is denoted $y=h$. The permittivities are $\eps_1$ and
$\eps_2$ and the electric fields are $E_1\hat{y}$ and $E_2\hat{y}$,
respectively.
Continuity of $\eps E$ across the interface together with $hE_1+(L-h)E_2=V$ gives us
$E_1=\eps_2E_0/(\eps_1+h\Delta\eps/L)$ and $E_2=\eps_1E_1/\eps_2$, where $E_0=V/L$ is the
average electric field and $\Delta\eps=\eps_2-\eps_1$ is the ``dielectric
contrast''. 
The electrostatic energy per volume of the
condenser is 
\begin{eqnarray}
\frac{F_{\rm es}(h)}{LS}=
-\frac12\frac{\eps_1\eps_2E_0^2}{\eps_1+h\Delta\eps/L}~.
\end{eqnarray}
As a result, if $\Delta\eps>0$, the system will reduce its energy if $h$ becomes as small
as possible, $h=0$. The maximum energy is when the interface is at $h=L$. The
electrostatic pressure on the interface is 
\begin{eqnarray}\label{p_es}
p_{\rm es}(h)=-\frac{1}{S}\frac{\partial F_{\rm es}}{\partial
h}=-\frac12\frac{\eps_1\eps_2\Delta\eps
E_0^2}{\left(\eps_1+h\Delta\eps/L\right)^2}~.
\end{eqnarray}
The sign reflects the fact that the electrostatic force tends to thin the
film if $\Delta\eps>0$.

Let us now go a step further in elucidating the role of dielectric interfaces by allowing
the flat interface to break into two parts of equal area with heights $h$ and $L-h$, such
that the volume of the fluids is conserved [Fig.~\ref{fig_liquid_para_perp}(b)]. For
simplicity we assume the initial interface to be at $y=L/2$. We ignore edge effects and
use the previous exercise for the electrostatic energy to get
\begin{eqnarray}\label{es_energy_broken}
\frac{F_{\rm
es}(h)}{LS}=-\frac12\frac{\eps_1\eps_2\bar{\eps}E_0^2}{\eps_1\eps_2+
(h/L)(1-h/L)(\Delta\eps)^2}~.
\end{eqnarray}
Here, $F_{\rm es}$ is symmetric around $h=L/2$, and the analysis shows us that the minimum
energy is achieved when $h=0$ ($h=L$ is equivalent). At this state, the fluid interface
perpendicular to the field has disappeared and was replaced by an interface parallel to
the external field [Fig.~\ref{fig_liquid_para_perp}(c)]. The difference between the
initial ($h=L/2$) and final ($h=0$) states is 
\begin{eqnarray}
\frac{\Delta F_{\rm es}}{LS}=\frac12 \frac{(\Delta\eps)^2}{\bar{\eps}}E_0^2~.
\end{eqnarray}
This important result shows us that (i) dielectric interfaces parallel to the
external field are electrostatically favored over perpendicular interfaces
and that (ii) the energy difference between the two cases scales like $E_0^2$
and  is proportional to $(\Delta\eps)^2$.
The deformation of a dielectric drop in electric field is in line with this
understanding -- by elongating along the field, the drop decreases the area of 
surface perpendicular to the field compared to the unperturbed sphere.

\subsection{The instability in pure dielectric liquids}

\begin{figure}[h!]
\includegraphics[scale=0.45,bb=25 550 528 775,clip]{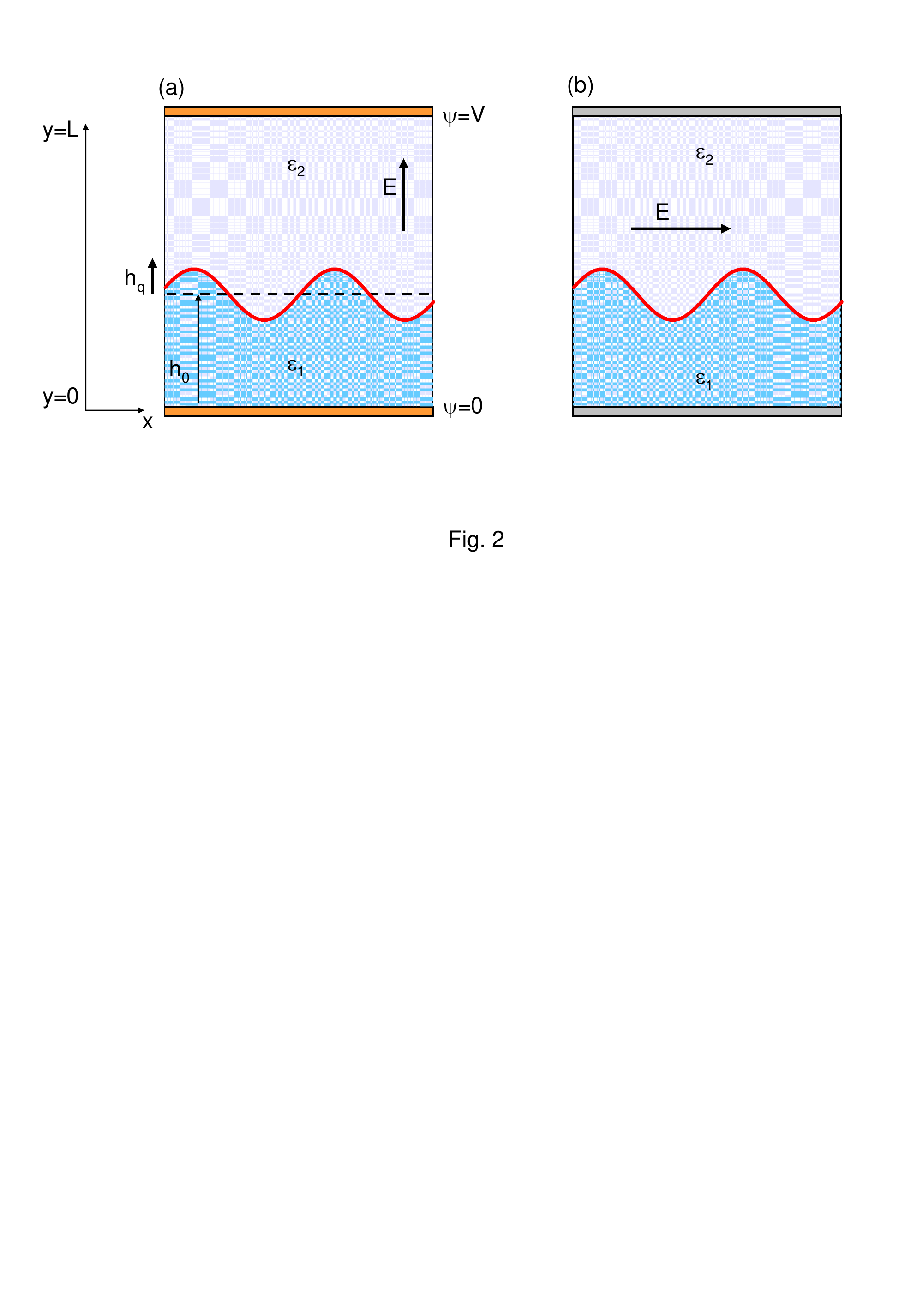}
\caption{Stabilization and destabilization by electric field. (a) Two liquids in
perpendicular electric field. $h_0$ is the unperturbed
interface location. Interface perturbations are unstable if their wavelength is long
enough. (b) A liquid film is stabilized in tangential electric field.
}
\label{fig_liquid_instab}
\end{figure}
The above derivation leads to the intuitive understanding that there is an electrostatic
free energy
penalty for dielectric interfaces perpendicular to the external field. Thus, the
perpendicular field
destabilizes the fluid interface initially parallel to the substrates, as found by 
\textcite{taylor1965} and \textcite{melcher1969} [Fig.~\ref{fig_liquid_instab}(a)].
Melcher also noted that an imposed electric field {\it stabilizes} an interface 
parallel to it \cite{melcher1968,onuki_physicaa1995}.
In the following we
present a simplified treatment of
the dynamical destabilization process following the lines
of  \textcite{hermin1999} and \textcite{russell_steiner2000}. 
The main aim is to find how the period and time constant of the pattern evolving
in the film depend on the surface tension and electric field.

Consider fluid confined between two parallel and flat electrode separated by a
distance $L$ and potential difference $V$, as is schematically depicted in Fig. 
\ref{fig_liquid_instab}(a).
The fluid has dielectric constants $\eps_1$
and unperturbed thicknesses $h_0$. We consider first the thin-film case, where the gap
contains a gas with dielectric constant $\eps_2$. Generalization to the bilayer case of
two viscous fluids is relatively easy.
As a simple approximation, the flow profile is assumed to be Poiseuille-like, and the
integrated current along the $y$ direction, $j(x)$, satisfies
\begin{equation}\label{poiseuille}
j(x)=-\frac{h^3}{3\eta}\frac{\partial p}{\partial x}~,
\end{equation} 
where $\eta$ is the liquid viscosity and $x$ is the direction parallel to the
interface.

There are three different contributions to the pressure on the film $p$: the
first one
comes from the interfacial tension $\gamma$ between the two fluids, $p_{\rm int}=-\gamma
h''(x)$. The second contribution is electrostatic, as given by Eq. (\ref{p_es}). For thin
enough films, van der Waals forces come into play, giving rise
to a disjoining pressure $p_{dis}=A/6h^3$, where $A$ is the Hamaker constant. For thin
liquid films, gravity may be neglected; however, gravity effects may easily be
incorporated \cite{onuki_physicaa1995}.

One may consider perturbations of the flat interface $h(x)=h_0+\delta h(x,t)$,
where $\delta h\ll h_0$.
The linear stability analysis will be restricted to the long
wavelength limit,
$\delta h'(x)\ll 1$. 
We may thus expand
$p_{\rm es}$ and $p_{\rm dis}$ to linear order in $\delta h/h_0$ to obtain 
\begin{equation}\label{p_film}
p(x)\simeq -\gamma \delta h''-\frac{A}{2h_0^4}\delta
h-\frac{\eps_1\eps_2(\Delta\eps)^2E_0^2}{L\left(\eps_1+h_0\Delta\eps /L\right)^3}\delta
h+{\rm const.}
\end{equation} 

This expression for the pressure, together with Eq. (\ref{poiseuille}), is
used in the continuity equation $\partial h/\partial t+\partial j/\partial x=0$.
We look at surface waves
of the form $\delta
h(x,t)=h_qe^{t/\tau}\cos(qx)$, where $\tau$ is the characteristic exponential
time for the
modulation with wave number $q$ and $h_q$ is the amplitude. 
The dispersion relation between $\tau$ and $q$
is readily obtained to be \cite{russell_steiner2001b}
\begin{equation}\label{tau}
\frac{1}{\tau}=\frac{\gamma
h_0^3}{3\eta}q^2\left(\xi_e^{-2}-q^2\right)~.
\end{equation} 
The generalized ``healing length'' is defined by the relation
\begin{equation}
\xi_e^{-2}=\frac{A}{2\gamma h_0^4}+\frac{
\eps_1\eps_2(\Delta\eps)^2E_0^2}{\gamma L \left(\eps_1+h_0\Delta\eps
/L\right)^3}~.
\end{equation} 
A positive $\tau$ means the modulation $\delta h(x,t)$ grows in time; negative $\tau$
shows exponential decay. Clearly, all $q$'s smaller than $\xi_e^{-1}$ are unstable.
$\tau$ is infinite when $q\to 0$ since liquid must then be transported to very long
distances; the opposite limit, $q\to\infty$, is also reasonable because very short
wavelengths are rapidly attenuated due to surface tension.

The fastest-growing mode $q^*$ is given by $q^*=\xi_e^{-1}/\sqrt{2}$.
Since $A\sim 10^{-20}$ J, films thicker than few nanometers 
are dominated by the electrostatic forces even at moderate field
strengths $E_0$. If the dispersive part can be neglected, we find that the wavelength of
the fastest-growing wave $\lambda^*$ is
\begin{equation}\label{lambda_star}
\lambda^*=2\pi/q^*\sim \gamma^{1/2}/\Delta\eps E_0~.
\end{equation} 
Thus, the most unstable wavelength can be reduced by increasing the dielectric contrast
$\Delta\eps$ or the field or by decreasing the surface
tension between the two liquids $\gamma$.

This normal-field instability also occurs in bilayers of two
viscous liquids. 
A straightforward generalization \cite{russell_steiner2001a} gives the same
expression as
in Eq. (\ref{tau}), only with the viscosity $\eta$ replaced by a rather complicated
function of the two liquid viscosities, $C(\eta_1,\eta_2)$. The time scale
of the dynamics is accelerated by as much as $~50$ times,
while $q^*$ and $\lambda^*$ stay
intact. The main advantage is the possibility to use liquid pairs with small
interfacial tension, thereby reducing $\lambda^*$.

\subsection{Experiments}

Initial experiments were conducted on thin polymer films above the glass transition
temperature, with electrode gap $L<1~\mu$m. In addition to the basic
understanding of interfacial phenomena in electric fields, experiments were motivated by
the possibility of a new lithography technique \cite{chou1999}. Indeed, amplification of
the most unstable mode leads to a film with a well-defined periodicity.
\begin{figure}[h!]
\includegraphics[scale=0.45,bb=65 390 450 765,clip]{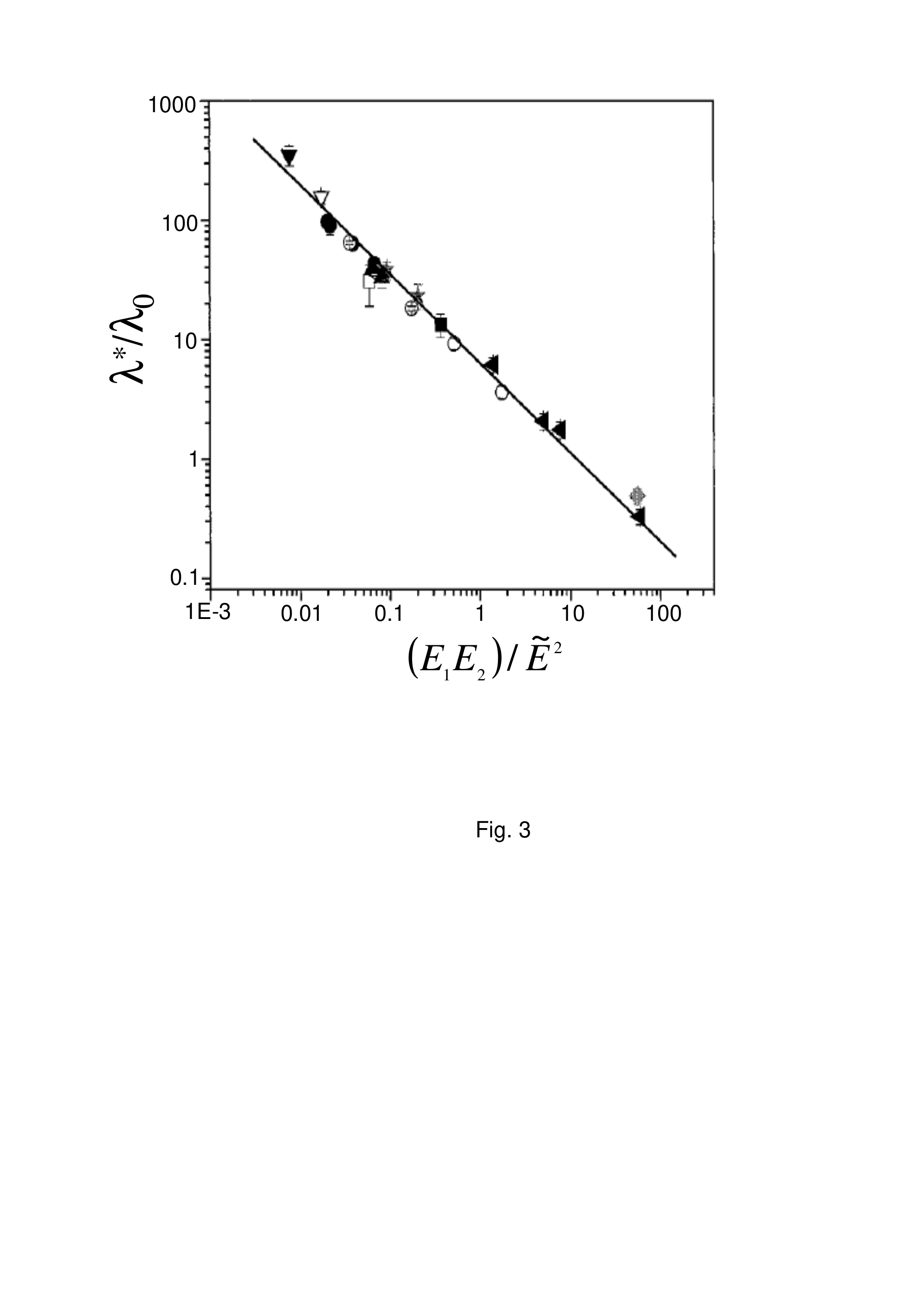}
\caption{Plot of the most unstable wavelength $\lambda^*$ vs reduced electric field for
various thin films. $E_1$ and $E_2$ are the electric fields in
the two liquids, $\lambda_0=V^2(\Delta\eps)^2/\gamma(\eps_1\eps_2)^{1/2}$,
and $\tilde{E}=V/\lambda_0$. The expected relation from Eq. (\ref{lambda_star}),
$\lambda^*/\lambda_0=2\pi(E_1E_2/\tilde{E}^2)^{-3/4}$, is the straight line with slope
of $-3/4$. 
Adapted with permission from \cite{russell_steiner2002}.
Copyright 2002 American Chemical society.
}
\label{fig_lambda_russell}
\end{figure}
Figure~\ref{fig_lambda_russell} shows a plot of $\lambda^*$ for a series of experiments
with different parameters: liquid thickness $h_0$, spacing $L$, voltage $V$, etc. 
The plot shows a good agreement with Eq. (\ref{lambda_star}) when the axes are
suitably defined \cite{russell_steiner2001a}. For a summary of data 
from several groups, the reader is referred to \textcite{russel2003}.

Hierarchical hexagonal structures have been obtained in sophisticated
experiments by the use of a trilayer, namely, a
liquid-liquid-air sandwich \cite{russell_steiner_nature_mater2003}. 
The resulting structures have two different periodicities,
depending on the dynamical process and on the relative volumes of the
components.

\vspace{0.5cm}

\subsubsection{Topographic electrodes}
\begin{figure}[h!]
\includegraphics[scale=0.65,bb=150 265 440 790,clip]{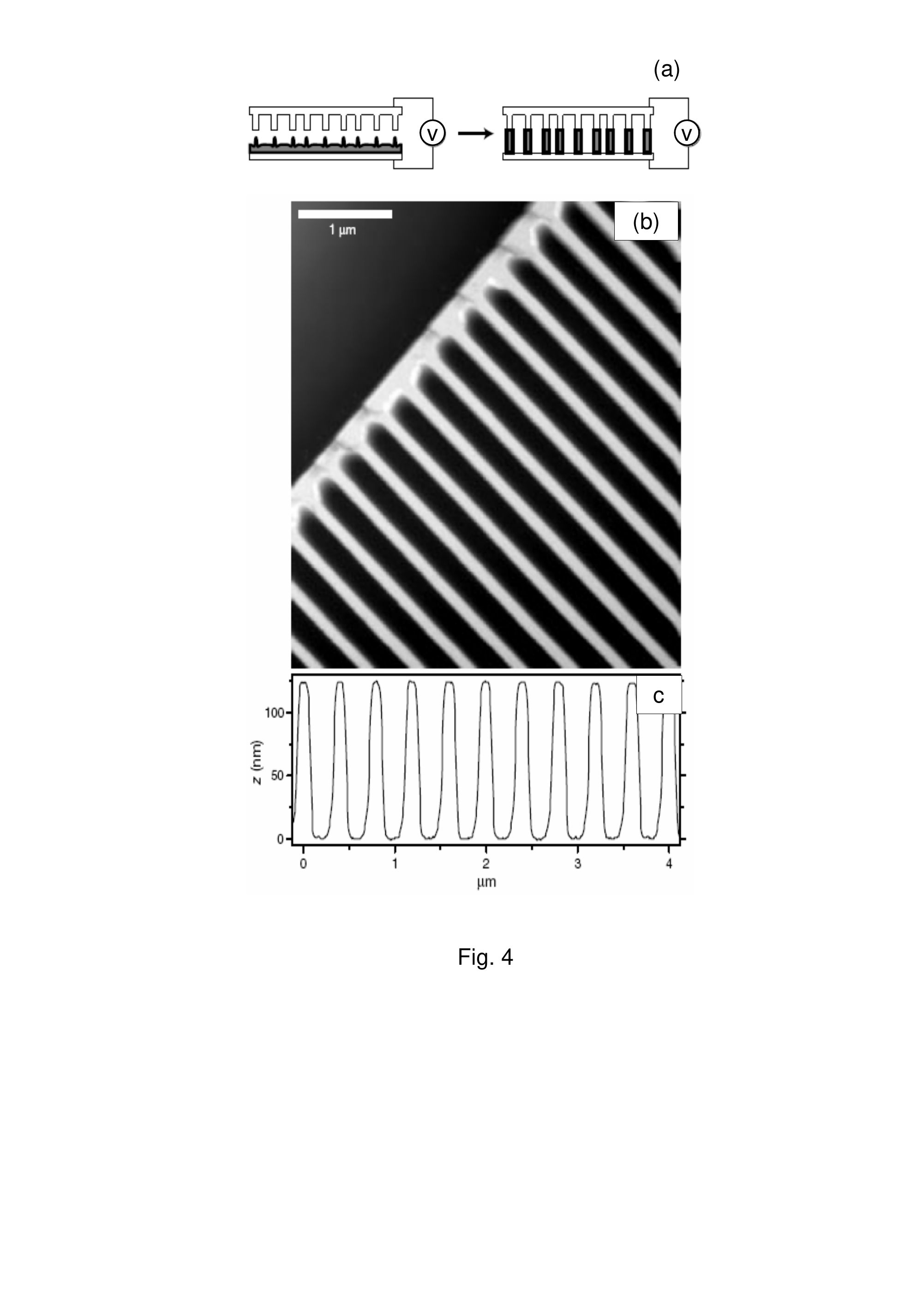}
\caption{Pattern transfer. 
Use of topographically patterned electrodes enables transfer of the top
electrode onto the polymer liquid, as is depicted schematically in (a). Tapping-mode 
atomic force microscopy 
images in (b) and (c) show the resulting topography of the polymer layer.
Adapted by permission from Macmillan Publishers Ltd: 
\cite{russell_steiner2000}.
}
\label{fig_top_elec}
\end{figure}

With smooth electrodes, the pattern period $\lambda^*$ [Eq. (\ref{lambda_star})] depends
on the electric field, interfacial tension, and dielectric contrast.
For
technologically motivated reasons, it may be beneficial to achieve different patterns. The
use of a topographically patterned electrode enables facile
and rapid duplication of a ``master mask'' onto the liquid film
\cite{russell_steiner2000} since the interfacial instability is amplified in places where
the electrode gap is smaller. Figure~\ref{fig_top_elec}
shows an implementation of this idea. Subsequent quench freezes the structure. 
\begin{figure}[h!]
\includegraphics[scale=0.45,bb=26 490 540 755,clip]{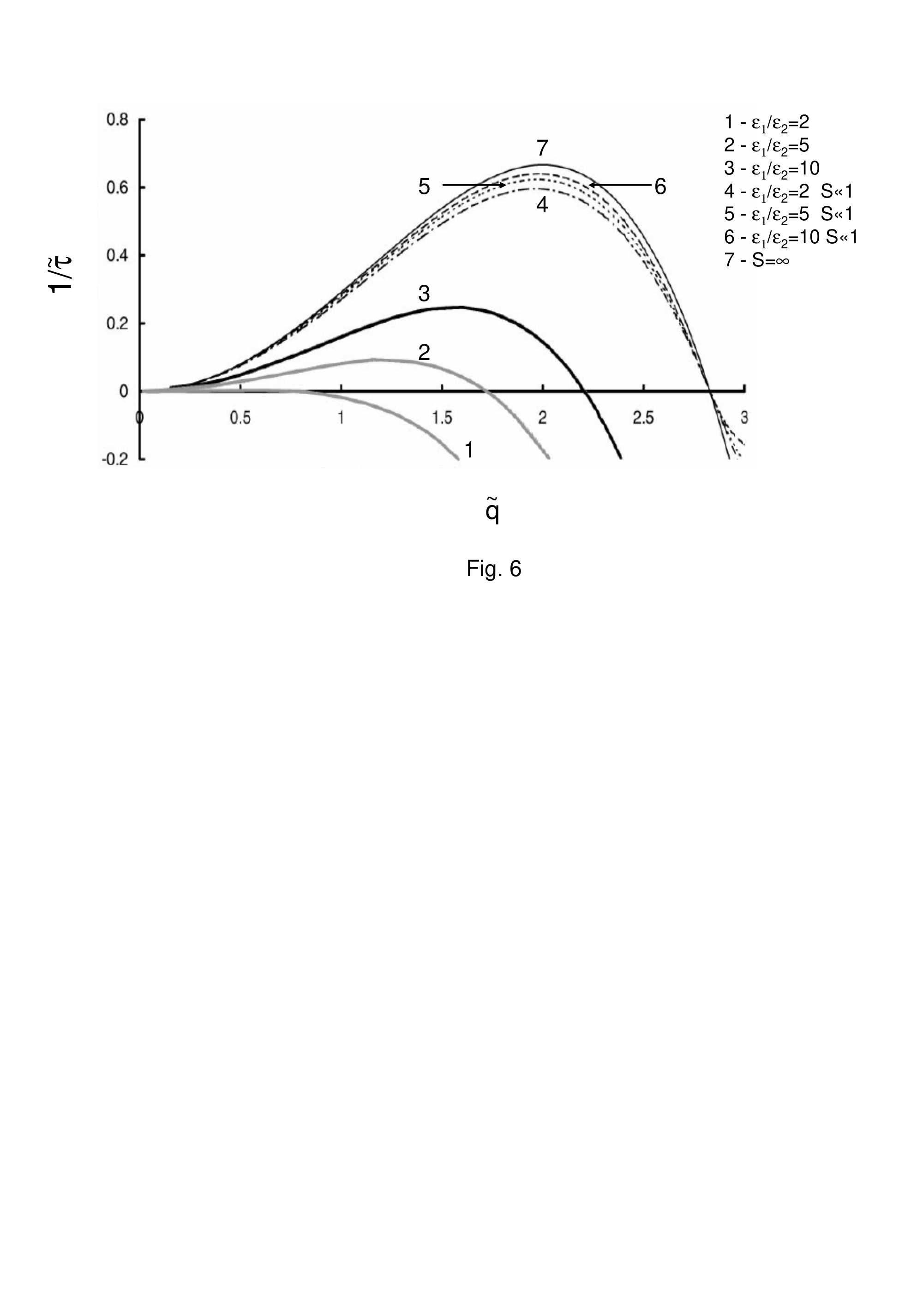}
\caption{Dispersion relation between wave number [scaled by $(\eps_2
V^2/\gamma L^3)^{1/2}$] and inverse growth time [scaled by $\eta_1\gamma L^3/\eps_2^2
V^4$]. Curves 1--3 are for pure dielectric liquids, Eq. (\ref{tau}). Curves 4--6
are from the ``leaky dielectric'' model with small dimensionless conductivity $S$.
The conductivity in curve 7 is infinite.
Reprinted from \cite{russel2002}, Copyright (2002), with permission from
Elsevier.
}
\label{fig_leaky_dielec}
\end{figure}
\subsubsection{Residual conductivity}
Work with polymer pairs gives smaller $\lambda^*$ because of the reduced 
surface tension, but at the same time
the dielectric contrast $\Delta\eps$ is diminished. How can this deficiency be overcome?
We must return to the role of residual conductivity in polymers. Conductivity
difference between two layered liquids leads to charge accumulations in
interfacial regions and to additional interfacial stress.

\textcite{russel2002} studied the application of Taylor's leaky dielectric model
to this system. Their analysis for slightly conductive polymer films yielded
growth
exponents and characteristic
wave numbers much larger than that of perfect dielectrics: even very small
conductivities decrease the growth exponent $\tau$ markedly and
decrease the fastest-growing wavelength $\lambda^*$ by a factor of $2$--$4$.
This is shown in Fig.~\ref{fig_leaky_dielec}, where $q$ and $\tau$ have been
appropriately scaled. The dimensionless conductivity $S$ is defined by
$S=\eta_2\gamma\sigma L^3/\eps_2^3V^3$, where $\sigma$ is the conductivity. The
numerical value of $S$ greatly varies and can be between $0.1$ and $10^{16}$.
Note that according to the model, small conductivity has a drastic effect on the
dispersion relation and that changing from $S\ll 1$ to $S>1$ does not change much.

\subsection{Near-critical fluids}

It is in order to briefly mention here near-critical fluids. The two fluids described
above may be a liquid in coexistence with its vapor phase or a binary mixture of two
partially miscible liquids. If one approaches the critical point from below, $T\to T_c$,
the surface tension and the dielectric contrast vanish like
$\gamma\simeq\gamma_0(1-T/T_c)^{2\nu}$ and
$\eps_2-\eps_1\propto(T_c-T)^\beta$, where 
$\nu\simeq 0.625$ and
$\beta\simeq 0.33$ are the exponents characterizing the correlation length and
the liquid-vapor density difference (zero-field magnetization in magnetic
systems). 
As \textcite{onuki_physicaa1995} pointed out,
this implies that $\lambda^*$ can be reduced upon approach to $T_c$ as
\begin{equation}
\lambda^*\sim \gamma^{1/2}/\Delta\eps\sim (T_c-T)^{0.295}~.
\end{equation} 

The relaxation dynamics of our simple model is influenced by the vanishing of
$\gamma$ as $T_c$ is approached and is different in the short- or
long-wavelength limits.
One should bear in mind, though, that the derivation used above
breaks down in this limit since we have used the assumption of sharp interfaces
between the liquids. The model is valid as long as the interfacial width between
the liquids $w\sim (T_c-T)^{-1/2}$ is much smaller than the unstable wavelength:
$w\ll \lambda^*$.

\subsection{Other related instabilities}

It is useful to mention several related instabilities originating from electric
fields.
\textcite{andelman_cr1986} considered a Langmuir monolayer of polar molecules.
These dipoles are constrained to the interface between two fluids, typically
water and air, pointing in a direction perpendicular to the interface. Since
parallel dipoles repel each other, their concentration may show modulations. On
writing the dipole surface density as $n(x)=n_0+n_q e^{iqx}$, they found the
electrostatic energy of the dipoles per unit area to be $F_{\rm
es}=-(1/2)|q|\eps_2/[\eps_1(\eps_1+\eps_2)]d^2n_q^2$. Here $d$ is the
electric dipole moment of individual molecules. The electrostatic energy thus
prefers a modulated state with infinitesimal period. 
The system is unconstrained
in the direction perpendicular to the two liquids, hence the $-|q|$ dependence. 
The competition between the long-range dipole-dipole repulsion and interfacial
tension between ``phases'', scaling as $q^2$, leads to an unstable mode
with finite wavelength \cite{mcconnel_pnas1984}. 
The simple analytical model, valid in the long-wavelength limit,
enabled them to construct a phase diagram of modulated phases with 
hexagonal and stripe symmetries \cite{andelman_jf1987}.
Similar behavior was found by
\textcite{garel_d1982} for thin uniaxial
ferromagnetic slabs subject to a perpendicular magnetic field even though the
physical
origin of the dipoles is different in the two cases. 
Another system of interest consists of a charged end-group polymer brush
placed inside a parallel-plate condenser. The
electrostatic energy per unit area associated with end-group
height undulation of the form $h(x)=h_0+h_q\cos(qx)$ can be written
to quadratic order in $h_q$  as
$F_{\rm es}=-\sigma^2/\eps |q|\cosh(qh_0)\cosh[q(L-h_0)]/\sinh(qL)h_q^2$. Here
$\sigma$ is the charge per unit area, $L$ is the surface separation and $\eps$ is the
dielectric constant of the uniform embedding medium. One retrieves the $F_{\rm es}\sim
-|q|$ dependence in the symmetric case where $h_0=L/2$ and in the limit $L\to
\infty$
\cite{elec_brush2008}. Again, the surface is unstable with respect to a finite
wavelength. This result essentially recapitulates the studies on the
instability appearing in $^3$He-$^4$He interface when it is charged with ions
\cite{leiderer_prl1979,leiderer_prl1984}.
The instability of Sec. \ref{interf_instab} is dynamical because
once the interface is parallel to the field, the field stabilizes it. Here,
however, the instability appears in equilibrium because due to its charge, the
interface is frustrated even if it is perpendicular to the substrate.

In a related work, \textcite{srolovitz2004} considered the stability of the
metal electrodes themselves to surface modulations. Two parallel metals
with potential difference $V$ and separated by an insulator are commonly found
in many situations, such as in micro-electro-mechanical-systems, microswitches, and close
to scanning transmission microscopy tips. They
used the same
long-wavelength approximation and obtained the electrostatic energy $F_{\rm
es}=-(1/8)|q|\eps E_0^2\coth(qL)h_q^2$, where $h_q$ is now the amplitude of electrode
surface modulation. When interfacial tension was added, they found that all
surface modes
with $q<q_c$ are unstable, where $q_c$ is given by $q_c=\eps
E_0^2\coth(q_cL)/4\gamma$.
However, the system is most unstable with respect to long wavelengths, that is, $q^*=0$.

The stability of a poorly conducting lipid membrane in aqueous environments in 
perpendicular electric field was studied as well \cite{sens_isambert2002}. The
accumulation of
charges on the opposite sides of the membrane is unbalanced if the membrane is not
completely flat. The destabilizing electric field acts like a negative surface tension,
tending to enlarge the membrane area. 
For a freely suspended membrane, a hydrodynamic
theory gives the fastest-growing $q$ mode $q^*\sim (E_0R)^{2/3}$, where $E_0$ is
the
external field and $R\simeq 10^5$ is the ratio between the membrane and solvent
resistivities. The corresponding wavelength is $\lambda^*=2\pi/q^*\simeq 0.5 \mu$m and
the growth rate is $\tau\simeq 10^5-10^6$ s$^{-1}$.

Lastly, we mention the normal-field instability in ferrofluids, an important
phenomenon discovered by \textcite{rosensweig_jfm1967}.
When the interface between a ferrofluid and a nonmagnetic fluid is
subject to a perpendicular magnetic 
field, the surface becomes unstable if the magnetization exceeds the
critical value, $M_c$.
The static pattern is hexagonal, and its period
$\lambda^*$ is given by $\lambda^*=2\pi(\gamma/g\Delta\rho)^{1/2}$
\cite{rosensweig_book,andelman_rosensweig_chapter}.

\subsection{Immiscible liquids in electric fields:
Electrowetting}\label{elec_wett}

Until now, we have described (i) how an initially flat liquid layer becomes unstable under
the 
influence of a perpendicular electric field and (ii) the deformation of a
liquid drop in
external fields. We now briefly describe an ``intermediate'' situation, that of a 
liquid drop placed on a solid flat substrate in electric field. Classical electrowetting 
describes the change in the wetting properties of two immiscible liquids due to the field 
\cite{mugele_wett_review2005}. This is a broad topic with many
important applications in microfuidics, lab-on-a-chip, etc.
\cite{ajdari_wett_review2004}. 

Consider first a dielectric drop embedded in a dielectric medium. 
For simplicity, we assume dc voltage and steady-state situation.
The elongation of dielectric drops not in contact with any
substrate, as considered by O'Konski and Thacher [see
Eq. (\ref{drop_elongation})], may lead
us to think that drop will elongate in the field's direction, thereby reducing the contact
area with the substrate and increasing the contact angle. The general shape change is
correct, but at the contact line this intuition fails: the contact angle $\theta$ stays
the same. The reason that $\theta$ is independent of $E_0$ is because the
electrostatic energy scales as the volume while the interfacial energies scale
as the area. Upon looking at
ever smaller regions close to the three-phase contact line, we thus find that
the electrostatic force becomes negligible compared to the interfacial forces.
The apparent contact angle, measured at a macroscopic scale, may be larger than
the field-free angle.

For a conducting drop, the situation is very different:
charge accumulation at a very thin layer at the substrate means the energy
contribution of the electric field becomes proportional to the surface and not to the
volume. This leads to a modification of
the interfacial properties and hence the apparent wetting angle changes from $\theta$ to
$\theta^*$. The main differences from Taylor's work on weakly conducting drops are 
(i) the existence of a solid surface and the localization of electric field at it 
and (ii) the embedding medium, usually a vapor, is nonconducting.

Lippmann's original work on electrolytes asserted that the
solid-vapor interfacial tension $\gamma_{\rm sv}$ is unaffected by the
potential, but
solid-liquid interfacial tension $\gamma_{\rm sl}$ is reduced by a value proportional 
to $V^2$ \cite{lippmann1875}. This reduction is due to the spontaneous creation
of an electric double layer at the substrate. However,
in the cases where the electrode is metal, electrolysis usually makes it
difficult to achieve high voltages. It is therefore beneficial to cover the
electrode with a dielectric insulator of thickness $d$ and permittivity $\eps$. 
In this situation,  the dielectric is responsible for the system's capacitance,
and the effective solid-liquid interfacial tension $\gamma_{\rm sl}^*$ is given
by \textcite{berge_cr1993},
\begin{eqnarray}
\gamma_{\rm sl}^*\simeq\gamma_{\rm sl}-\frac{\eps}{2d}V^2~.
\end{eqnarray}

The apparent contact angle $\theta^*$ is found from 
substitution of $\gamma_{\rm sl}^*$ in Young's equation and is given by
\begin{eqnarray}
\cos\theta^*=\cos\theta +\frac{\eps (V-V_0)^2}{2\gamma d}~.
\end{eqnarray}
The $V_0$ term accounts for spontaneous charging -- it is
common that a solid in contact with an electrolyte inherits a net charge by ion
adsorption
or ionization of covalently bound groups. For instance, common glass near water
ionizes to make SiO$^-$ and releases a proton.
The above supposes a wedge-shaped interface and is correct on a
macroscopic scale. On a scale smaller than thickness of the dielectric layer
$d$, the
curvature of the liquid-vapor interface is not fixed.

In the absence of a disjoining pressure, the interface shape
is governed by the generalized Laplace's equation,
\begin{equation}
\gamma\kappa({\bf r})-p_{\rm es}({\bf r})={\rm const.}
\end{equation} 
Here $\kappa$ is the local curvature and 
$p_{\rm es}\propto V^2$ is the electrostatic pressure, calculated globally for the drop
\cite{mugele_wett_review2005}. At the mesoscopic scale, a numerical procedure
assuming a circular contact line showed
that the electric field diverges weakly at the three-phase contact line, and the
slope approaches Young's angle, $\theta$, at the contact line 
\cite{hermin_mugele_prl2003,mugele_epl2006}. This theoretical finding has been
verified experimentally recently \cite{mugele_jphys_condens_matter2007}.
Note that the drop's shape needs not stay circular, and a static instability of
the contact line was observed in high voltages \cite{mugele_hermin_apl2002}. 
In contrast to the
instability discussed throughout Sec.~\ref{interf_instab}, here the 
complex interfacial shapes are static due to a balance between Laplace and
electrostatic pressures.

The influence of time-varying ac fields depends on the field's frequency
$\omega$: for a
liquid drop of dielectric constant $\eps$ and conductivity $\sigma$, the drop
behaves as in dc field if it is in the quasistatic regime, that is, 
when $\omega\ll \omega_c$ \cite{LL_electrodynamics,mugele_wett_review2005},
where
\begin{equation}\label{omega_c}
w_c= \sigma/\eps~.
\end{equation} 
In the opposite regime, $\omega \gg \omega_c$, the mobile ions do not have
enough time to response to the field before it changes sign. Electric double layer is not
created, and therefore the field acts throughout the whole drop volume. The electric field
thus exerts a body volume force again just like for static dc fields in pure
dielectrics. 
Experimentally, demineralized water have $\sigma\simeq 4\times 10^{-6}$
S m$^{-1}$ and
therefore at frequencies $\omega>10^4$ s$^{-1}$ the behavior is similar to the dielectric
case. The frequency $\omega_c$ appears also as an important measure of the
influence of ions on orientation of block copolymers by electric fields (see 
Sec. \ref{mobile_ions}).

As a final comment, we stress that all of the above is not valid close to a
critical point: if $T-T_c$ is small enough, the correlation length diverges, and
the interfacial
tensions $\gamma$, $\gamma_{\rm sv}$, and $\gamma_{\rm sl}$ become dependent on
the field
distribution and droplet shape.

\section{Block-Copolymer Orientation in Uniform Electric Fields}
\label{sect_bcp}

The preceding section described several examples of deformation of an interface
between two immiscible liquids or between a liquid in coexistence with its
vapor. In
both cases the two phases are macroscopic, separated by a thin interfacial
layer. An
interesting question regards the effect of electric field on complex materials which 
self-assemble into ordered structures with typical lengths on the mesoscopic scale. We
will concentrate on one such material -- block copolymers (BCPs). These consist of two or
more
chemically distinct polymer species connected by a covalent bond. Their self-assembly
results from a competition between enthalpic interactions and chain stretching
of entropic origin \cite{fred_physics_today1999}. Block copolymers have
attracted considerable
research in the past years because many of their interesting mechanical,
electrical,
rheological, and other properties can easily be fine tuned for optimum performance
in nanotechnological applications \cite{thomas_polymer2003}.

The phase behavior of diblock copolymers,
made up of two polymers A and B, is governed by two parameters: $\phi_0$, the volume
fraction of the A polymer ($0<\phi_0<1$) and $N\chi$. $\chi$ is the so-called Flory
parameter and is approximately inversely proportional to the temperature and
$N$ is the
polymerization index. In symmetric diblock copolymers, the A and B parts of the polymer
have the same length, and $\phi_0=1/2$. If the temperature is high enough (small
$N\chi$), the melt is found in a disordered phase. Cooling below the so-called
order-disorder temperature (ODT) leads to the appearance of a lamellar phase.
This is a one-dimensional and periodic phase of A- and B-rich domains of width
$d$ comparable to the polymer radius of gyration. 
At fixed temperature, further increase in the length asymmetry between the A
and B blocks (larger $|\phi_0-1/2|$) increases the spontaneous curvature: the
polymers adopt a configuration where the shorter block is confined to the inside
of cylinders arranged on a hexagonal lattice. Even larger increase in
$|\phi_0-1/2|$ leads to the formation of spheres arranged on a bcc phase. Other
more complicated ordered phases appear in block copolymers and even in simple
diblocks (e. g., gyroid phase), but we ignore them here. The region in the phase
diagram in the $(\phi_0,N\chi)$ plane close to the ODT point, given by
$N\chi\simeq 10.5$, is called the weak-segregation regime. In this region,
domain
spacing $d$ scales like $N^{1/2}$. The region farther below the ODT point,
$N\chi\gg 10.5$, is called the strong-segregation regime, where the A and B
blocks are highly segregated from each other, and $d\sim N^{2/3}$.

\subsection{Orientation mechanism}

The orientation of block copolymers and indeed many other soft-matter phases can be
influenced by several external fields. Examples are shear flow, van der Waals or
chemical interaction with surfaces, confinement by walls, etc.
At first inspection, it seems that electric fields are very weak since the
electrostatic energy stored in a molecular volume $v_0$, $v_0\eps E^2$, is much
smaller than the thermal energy $k_BT$. However, experiments have shown that
this estimate is too simplistic and $v_0$ is not the relevant volume.
The first experiments on block copolymers in electric fields were carried out by
\textcite{ah_mm1991}. In a series of papers they developed the key concepts for
alignment and stability of lamellar phases in electric fields \cite{ah_mm1993,ah_mm1994}.
In the first set of experiments, polystyrene-polymethylmethacrylate (PS/PMMA)
was anionically synthesized.
Samples were heated above the ODT point, held for about 10 min, and then
cooled below the ODT in the weak-segregation region. In the absence of an electric
field, a small-angle x-ray scattering (SAXS) reveals a characteristic ring,
corresponding to a repeat period of $d=23$ nm. 
Samples under electric field, however, exhibit strong anisotropy, as is manifest
in the
clear peaks appearing in the direction perpendicular to the electric field.
\begin{figure}[h!]
\includegraphics[scale=0.6,bb=100 430 495 775,clip]{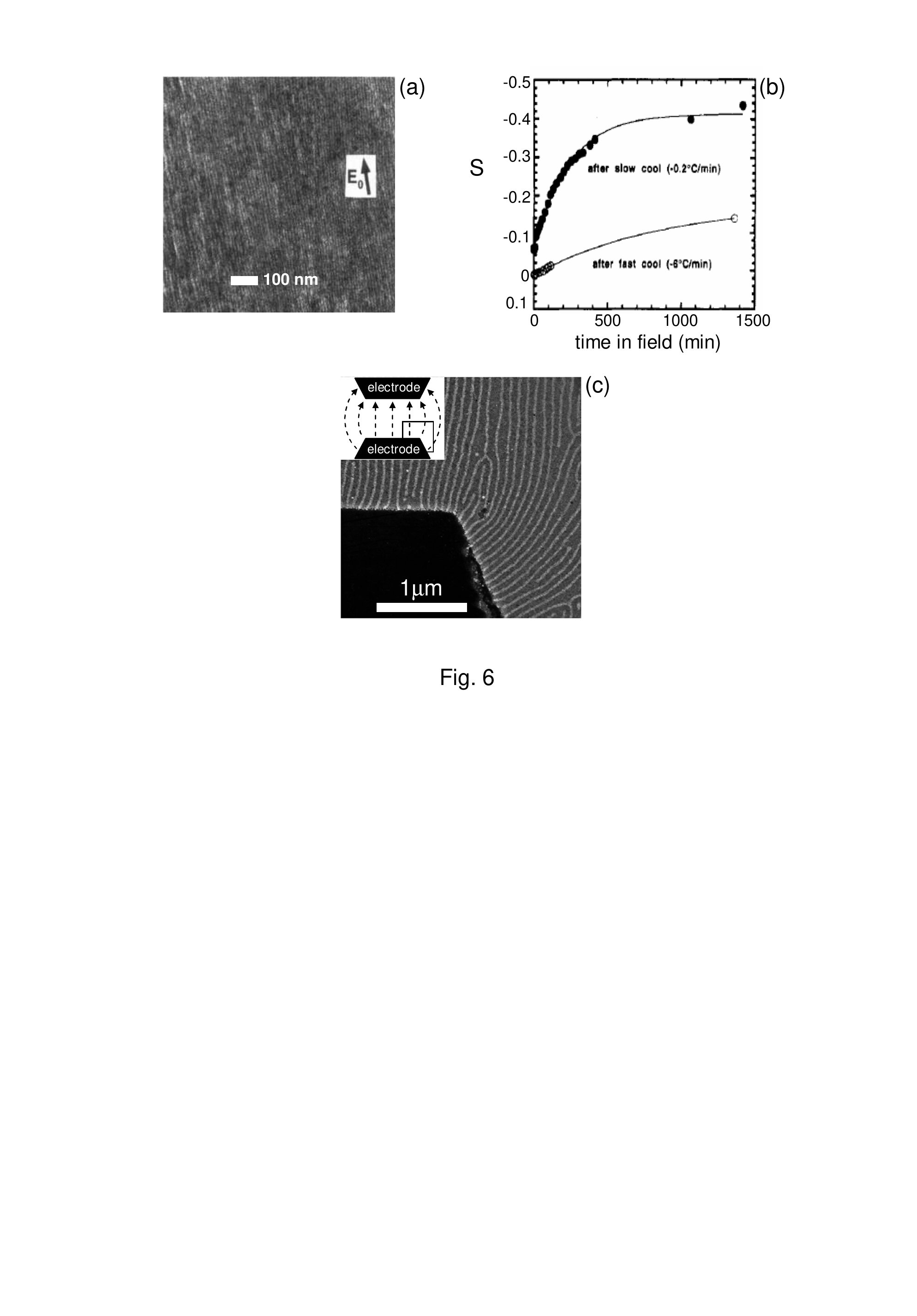}
\caption{Orientation of PS/PMMA block copolymer lamellae. 
(a) The bulk experiments of \textcite{ah_mm1993} show near-perfect alignment of
lamellae in the direction parallel to the external field. (b) Time evolution of
the orientational order parameter $S$ for two samples with different cooling
rates as measured by birefringence experiments. Solid line is an exponential
fit. 
Adapted with permission from \cite{ah_mm1993}. Copyright
2002 American Chemical society.
(c) TEM closeup of cylindrical diblock copolymers near an electrode.
Inset is schematic illustration of the full setup. Dashed arrows are field
lines and box indicates area of image. Note how
cylinders curve and follow the field lines near the corner. 
From \cite{russell_sci1996}. Reprinted with permission from AAAS.
}
\label{fig_bcp_orient}
\end{figure}

Figure~\ref{fig_bcp_orient}(a) is a transmission electron micrograph
(TEM) of lamellae
oriented in electric field. In order to quantify the alignment of ordered phases,
\textcite{ah_mm1993} used the quantity $S$ defined as follows.
Denote by $\phi({\bf r})$ the local A-polymer composition ($0<\phi<1$).
In the disordered phase, $\phi$ is equal to its average value $\phi_0$. One can
write $\phi({\bf r})$ as a sum of plane waves of the form
\begin{eqnarray}\label{op_fourier}
\phi({\bf r})&=&\phi_0+\varphi({\bf r})~,\nn\\
\vphi({\bf r})&=&\sum_{{\bf q}\neq 0}\vphi_{\bf q}e^{i{\bf q}\cdot{\bf r}}~.
\end{eqnarray} 
For example, close to the ODT point, a lamellar phase can be described by $\phi({\bf
r})=\phi_0+\vphi_L\cos(qx)$, where $\vphi_L$ is an amplitude and the normal to the
lamellae
is
chosen in the $x$ direction. Similarly, hexagonal and cubic phases contain three and six
modes, respectively.
For a lamellar phase, the average orientation $S$ is then defined by
\begin{equation}\label{S_op}
S=\frac32\langle (\hat{q}\cdot \hat{E}_0)^2\rangle-\frac12~.
\end{equation} 
In nematic liquid crystals, $S$ is commonly used to quantify order.
In the beginning of the process, there are many random grains in different
directions
not correlated with the field, the sample is macroscopically disordered, and therefore
$S=0$. As orientation of lamellar grains proceeds, ${\bf q}$ turns until it
becomes perpendicular to ${\bf E}_0$ for reasons which will become clear below. Perfect
orientation has $S=-1/2$. Figure~\ref{fig_bcp_orient}(b) is a time-evolution
plot of $S(t)$.
The evolution is slow because the samples are very viscous. The sample with slow cooling
rate equilibrates faster to the electrostatically preferred orientation. Figure
\ref{fig_bcp_orient}(c) is a TEM image taken by \textcite{russell_sci1996} close to the
electrodes. Clearly, lamellae seem to follow the curved field lines.

\subsection{Experiments}

The recent decade has witnessed an explosion of work on electric-field
effects in block copolymers. Experiments can be roughly divided to two types:
those carried out in a block copolymer melt and those in a solution. In the
first case, pure block copolymer melt is subject to an external electric field.
Since the melt is very viscous, the sample is heated to elevated temperatures 
and subsequently annealed under electric field.  SAXS, TEM, and small-angle
neutron scattering are used to characterize the copolymer
structures.
Annealing times are very long, and therefore experiments in copolymer melts
intrinsically probe static phases. On the other hand, when copolymers are
dissolved in a solvent, the viscosity is not high and orientation kinetics in
electric field can be recorded. 

\subsubsection{Statics}

The experiments of \textcite{ah_mm1994} showed that block copolymer structures can be
oriented in the bulk. This work has stimulated further attempts to orient thin
film structures. The advantage in using thin films is their possible
technological applications and the reduction in required voltage following the reduction
in size. The disadvantage in size reduction is that the electric field, acting throughout
the sample volume, needs to overcome increasingly more dominant interfacial energies
acting between the copolymers and the bounding surfaces.

The experiment in Fig.~\ref{fig_bcp_orient}(c) and others \cite{russell_mm1998}
showed that BCP domains can
be oriented laterally along an external field even in a thin film. 
Consider a BCP melt sandwiched between two flat and parallel surfaces at
distance $L$ from
each other and potential difference $V$. One would naively expect that there is one
transition field, below which BCP lamellae (or cylinders) are parallel to the
substrate and above which domains are perpendicular to it.
The electric field acts throughout the sample
volume, and therefore the 
order of magnitude of field strength
required for effective orientation is given by $\eps E^2\sim \gamma/L$, where
$\gamma$ and $\eps$ are the typical
difference between the interfacial tensions of the polymers with the substrate 
and dielectric constant. In most experimental systems $\gamma\sim 10$-$100$mN,
$L\sim 10$-$100~\mu$m, and $\eps\sim 5\eps_0$, and we therefore find $E\sim 10^3$-$10^7$
V/m.
The larger value of these estimates is quite a large field, but it is nonetheless
considerably
smaller than the dielectric breakdown threshold, $E\sim 10^8$ V/m.
\begin{figure}[h!]
\includegraphics[scale=0.7,bb=160 575 395 765,clip]{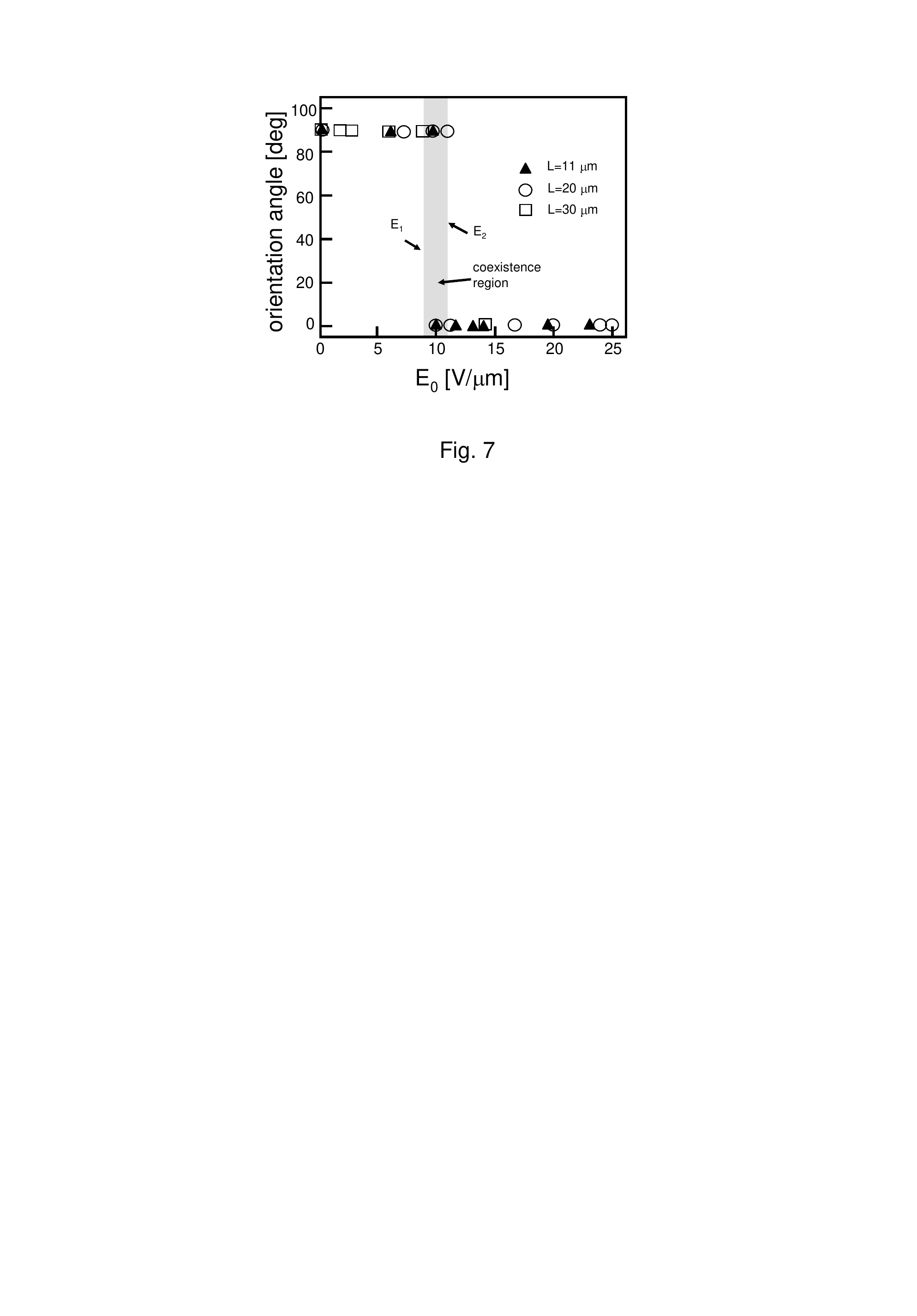}
\caption{Summary of sample orientation with respect to the field.
Orientation angle is the
angle between cylinders' axis and external field for three different film
thicknesses $L$.
Shaded area is a coexistence ``mixed'' region, where both orientations are observed.
Critical fields $E_1$ and $E_2$ 
mark the transition fields between the three possible states (cylinders parallel or
perpendicular to the surface and a mixed state). 
Adapted with permission from \cite{russell_mm2000}. Copyright
2002 American Chemical society.
}
\label{fig_exp_scat_coexist}
\end{figure}

Subsequent work on BCP thin films verified that electric field can indeed orient lamellae
and cylinders from a direction parallel to the substrate to a direction perpendicular to
it (and parallel to the field) \cite{russell_mm2000}. 
Anionically synthesized samples of PS/PMMA in the disordered (high temperature) state were
exposed to electric fields of $1$--$25$ V/$\mu$m. One aluminum electrode was in direct
contact with the copolymers, while the second one was insulated from the copolymers by a
Kapton sheet. The sample was then cooled down to the cylindrical phase and held for 14 h,
and finally cooled to room temperature. SAXS experiments, shown on Fig.
\ref{fig_exp_scat_coexist}, revealed the following
general picture: surprisingly, there are two critical fields $E_1$ and $E_2$. If $E<E_1$, 
cylinders are parallel to the substrate, while above $E_1$ they are destabilized. 
If $E_1<E<E_2$, there is a mixed state, showing characteristics of both parallel and
perpendicular cylinders. If $E>E_2$, the sample is fully oriented perpendicular to the
substrate and parallel to the external field.

As mentioned above, due to the high polymer viscosity, experiments in the melt
can be performed with polymers of limited molecular weight. Typically, large
electric fields are required in order to overcome defects that
persist in the sample and to accelerate
the slow dynamical process. A different approach was presented by
\textcite{boker_mm2002},
whereby copolymers are dissolved in a solvent. 
The use of a polymer solvent allows additional experimental flexibility: for
example, a selective solvent for one of the polymers may, in principle,
effectively increase the dielectric contrast $\Delta\eps$, and it also
enables use of high molecular weight and
branched polymers. In addition, because the solution is less viscous than the melt,
the full orientation kinetics can be recorded in real time.
However, a usual deficiency of this approach is the nonselectivity of common solvents: 
as the solvent dissolves in the respective polymer blocks, the effective
dielectric contrast $\Delta\eps$ between domains is diminished.

Bulk samples of polystyrene-poly(2hydroxyethyl methacrylate)-polymethylmethacrylate
(PS/PHEMA/PMMA) with number-average molecular weight M$_n$ of 82\,000 g/mol were dissolved
in chloroform and put in a cylindrical capacitor with average electric field $E_0=1.8$
V/$\mu$m. 
The solvent evaporated in a controlled manner and under the influence of external
field,
and the resulting samples were characterized by SAXS, TEM, and differential scanning
calorimetry.
These techniques show that the PHEMA
block is miscible with the PMMA block and that the triblock copolymer actually
behaves
very similar to a PS/PMMA diblock of enhanced dielectric contrast.
The experiments showed that block-copolymer solutions can be efficiently oriented
parallel to the external field, overcoming interfacial interactions. In addition, they
also opened a window for quantitative consideration of defect statistics and, more
importantly, the dynamics of grain orientation and defect annihilation.

\subsubsection{Dynamics}

As is explained above, one of the advantages of orientation in polymer
solutions is the possibility to track orientation dynamics. In order to record
dynamics, a high-flux scattering source is required because the exposure time is
limited as compared to static experiments. Another important consideration is
the copolymer volume fraction in solution: less copolymer not only reduces the viscosity 
but also reduces the effective dielectric contrast $\Delta\eps$. The electric
field needs to be high enough so that orientation terminates before solvent
evaporation completes and the structure becomes virtually immobile.
\textcite{boker_mm2003} used lamellar-forming polystyrene-polyisoprene
(PS/PI) block copolymer 
dissolved in toluene. The samples were exposed to synchrotron SAXS of high beam
energy and photon flux during exposure to electric field.
The synchrotron beam direction was perpendicular to the
electric field's direction. The measured average sample orientation, as
indicated by $S(t)$ of Eq. (\ref{S_op}), showed a similar qualitative behavior
as in Fig.~\ref{fig_bcp_orient}(b). But there is a big difference -- the
dynamics are 20-fold faster. The relaxation of $S(t)$ is characterized by a
single exponential with time constant $\tau$. As expected, $\tau$ increases with
increasing copolymer content and decreasing temperature.

Using in situ SAXS experiments, \textcite{boker_mm2003} identified 
two mechanisms for orientation of ordered phases: close to the ODT point, 
the polymer solution 
undergoes a transition between only two orientations (from perpendicular to
parallel to
${\bf E}_0$), with no intermediate orientations. Grains of lamellae in the favorable
direction grow on the expense of unfavorable grains, resulting in grain boundary
migration. However, far from the ODT point
(lower temperatures), the scattering pattern has all the intermediate
grain
orientations and thus reflects continuous rotation of grains to the 
favorable
direction. The optimum grain orientation is not fully achieved since the driving 
force
for orientation, torque, is negligibly small at long times.

\begin{figure}[h!]
\includegraphics[scale=0.5,bb=90 465 470 765,clip]{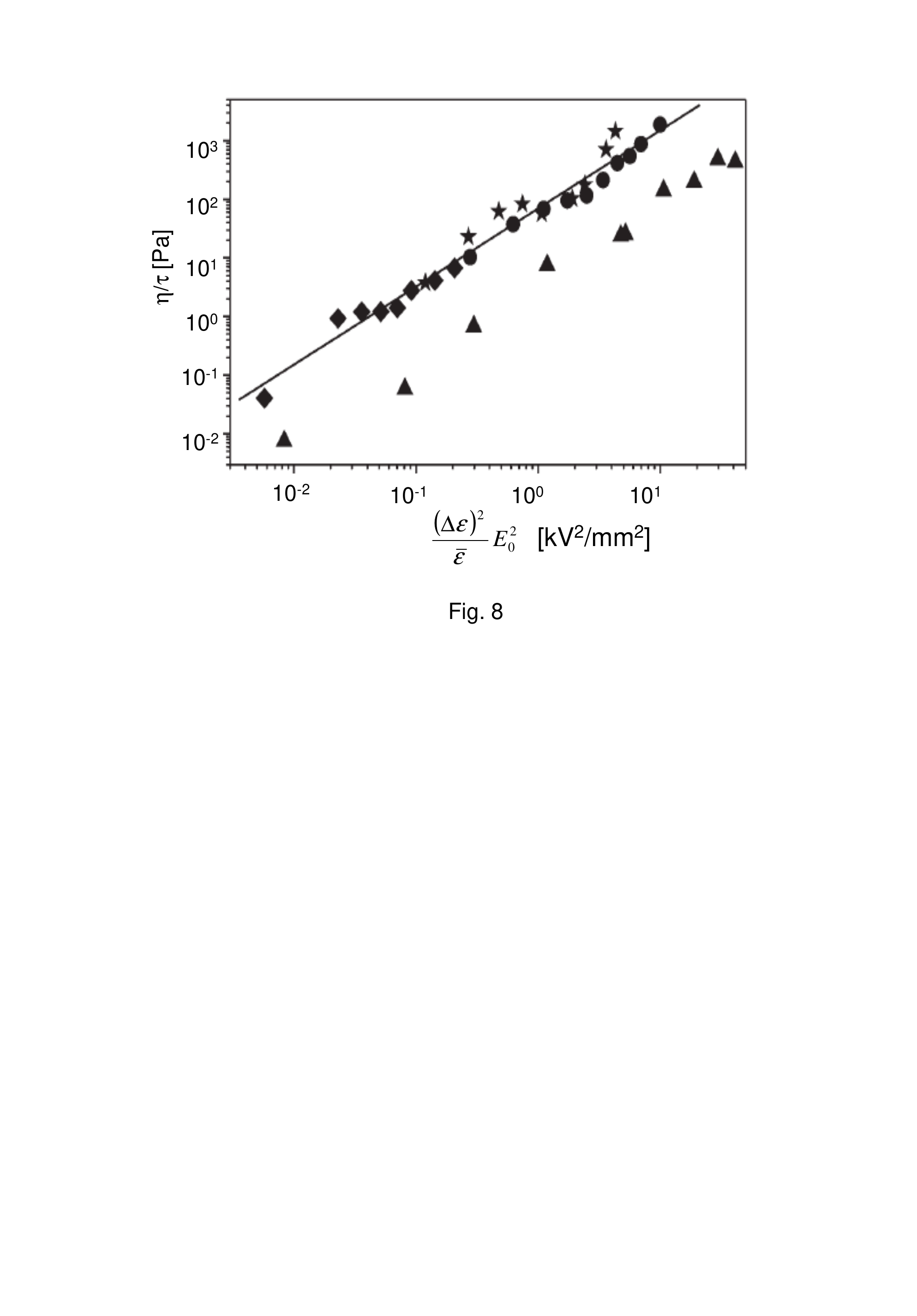}
\caption{Inverse exponential relaxation time $\eta/\tau$ against $E_0^2$
from four
copolymer solutions as obtained by SAXS. When properly scaled, most data sets fall on a
single universal curve. From \textcite{boker_soft_matter2007}.
Reproduced by permission of The Royal Society of Chemistry.
}
\label{fig_scaling_field}
\end{figure}

What is the nature of the ``driving force''? Figure~\ref{fig_scaling_field} shows the
collection of results from numerous systems with different copolymer solutions.
For all four copolymer solutions, the inverse exponential relaxation time 
$\tau^{-1}$ scales linearly with $E_0^{2.7}$, with scaling prefactor depending on the
specific solution. The exponent, being $2.7$ and not $2$, is attributed to alignment
process governed by an energy barrier; this is a feasible explanation in the
weak-segregation regime, where grain orientation is dominated by nucleation and growth
\cite{boker_soft_matter2007}.
A simple scaling form exists validating the importance of the dielectric contrast
$\Delta\eps$: when the data are presented for $\eta/\tau$ against
$(\Delta\eps)^2E_0^2/\bar{\eps}$, where $\eta$ is the solution viscosity, all data except
one collapse onto a universal curve. 

A general theory of orientation of ordered phases is presented below based on a
coarse-grained simplified approach.

\subsection{Theory}

The underlying physical mechanism of copolymer orientation, sometimes referred
to as the ``dielectric interfaces,'' is described below.
This mechanism relies on dielectric contrast between the
different material domains. The electric field favors one sample orientation
over another, the energy difference being proportional to $(\Delta\eps)^2E_0^2$.
Intuitively, we may say that there is a free energy penalty for having
dielectric interfaces perpendicular to the field's direction. This is true in
both the weak- and strong-segregation regimes, but the analytical theory is
different in the two cases. 

\subsubsection{Electrostatics of strongly segregated lamellae}
\begin{figure}[h!]
\includegraphics[scale=0.4,bb=20 575 555 745,clip]{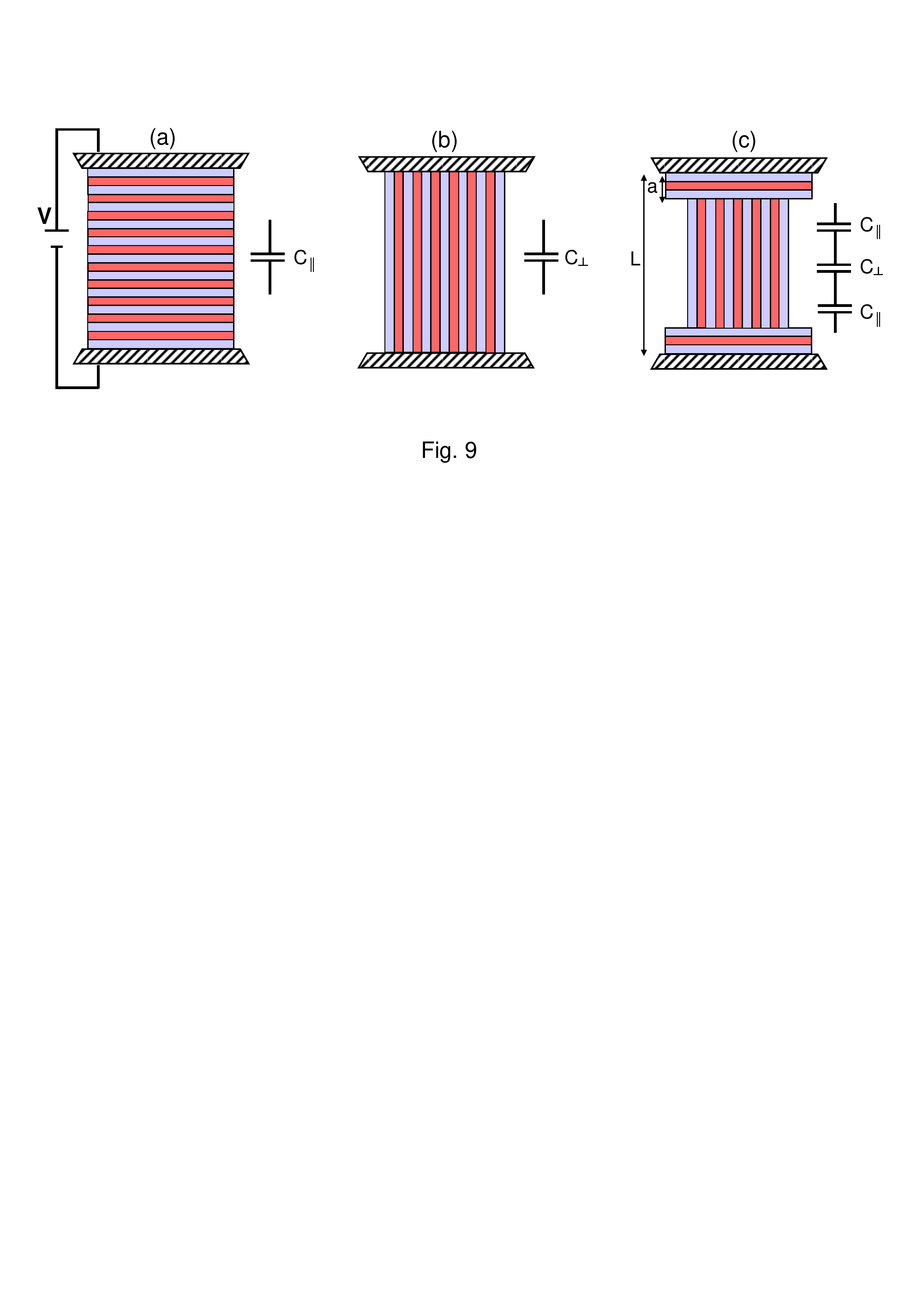}
\caption{Capacitor analogy for three competing lamellar structures confined in a
condenser. (a) Parallel and (b) perpendicular layering and (c) mixed morphology, 
consisting of two parallel layers of thickness $a$ at the electrodes and perpendicular
lamellae throughout the rest of the film. The equivalent electrical circuits are
illustrated on the left.}
\label{fig_para_perp_mix}
\end{figure}

All interesting effects reviewed here are due to dielectric anisotropy. One obvious
anisotropy in $\eps$ is due to the mesoscopic order, in which $\eps$ is a function of
${\bf r}$. Another contribution may come from the inherent anisotropy of $\eps$ in the
presence of electric field. Indeed, since polymers are large objects with many
chemically differing regions, $\eps$ in general is a tensor, and therefore the dielectric
response is anisotropic in nature. Here we ignore the tensorial nature of
$\eps$ since it is not observed in most experiments. The 
interested reader may turn to the theoretical description of block copolymers
with anisotropic $\eps$ given by \textcite{guro_mm1994, guro_prl1995}.

Strongly segregated lamellae are characterized by a square-wave composition profile
$\phi$. For the electrostatic calculation it is therefore reasonable to take lamellae as
being composed of pure A and pure
B polymers, with dielectric constants $\eps_1$ and $\eps_2$, respectively, separated by a
sharp interface. The capacitor model allows the
calculation of electrostatic energy of competing
orientations \cite{per_williams_mm1999,TA_mm2002}.
Consider a stack of lamellae parallel to the electrodes separated by distance $L$ and
potential difference $V$, as illustrated in Fig.~\ref{fig_para_perp_mix}(a). Such
parallel lamellae may be stabilized by preferential interactions with the
substrate. For
symmetric lamellae, we use the continuity of the displacement field $D=\eps E$
across dielectric interfaces and the total potential drop and find the
electrostatic energy
per unit area to be $F_{\rm es}^\parallel=-\frac12 C_\parallel V^2$, where the
capacitance per unit area $C_\parallel$ is given by
\begin{equation}
C_\parallel=\frac{\eps_1\eps_2}{\bar{\eps}}\frac{1}{L}~.
\end{equation} 
Here $\bar{\eps}=(\eps_1+\eps_2)/2$ is the average dielectric constant. Similarly, for the
perpendicular arrangement [Fig.~\ref{fig_para_perp_mix}(b)] we have $F_{\rm
es}^\perp=-\frac12 C_\perp V^2$ and 
\begin{equation}
C_\perp=\frac{\bar{\eps}}{L}~.
\end{equation} 

Since $C_\perp$ is larger than $C_\parallel$, the electrostatic energy $F_{\rm es}$ is
always lower for the perpendicular lamellae, with the difference between the two
orientations being proportional to $(\Delta\eps)^2$. Which of the two phases,
parallel or
perpendicular lamellae, is thermodynamically stable that is a matter of a competition
between
the interfacial energies, scaling like the substrate area, against the
electrostatic energy,
scaling like the sample volume. It is clear that there is a critical voltage
above
which perpendicular lamellae will be preferred over parallel ones. This voltage depends
on $\Delta\eps$, $L$, and the interfacial energies.

The two structures above are not the only conceivable ones. More complex phases,
such as the one depicted in Fig.~\ref{fig_para_perp_mix}(c) may be possible. 
This mixed morphology presents
an interesting compromise: the system keeps few parallel layers at the
substrates, thus optimizing interfacial interactions, but has lamellae oriented
in the field's direction in the rest of space, thus minimizing electrostatic
energy throughout most of the film. However, there is an energetic penalty per
unit area of the film, $\gamma_{_T}$, for the creation of a ``T-junction''
defect.

The capacitor model allows us to calculate
$F_{\rm es}$ for the mixed state as well.
The equivalent electrical circuit is that of two
``parallel'' capacitors and one ``perpendicular'' one connected in series. Ignoring
edge or ``fringe'' effects, we may write the capacitance of the mixed morphology
by
\begin{equation}
\frac{1}{C_m}=\frac{2a}{L}\frac{1}{C_\parallel}+\frac{L-2a}{L}\frac{1}{C_\perp}
~.
\end{equation} 
The factors in front of the inverse capacitances on the right hand side are necessary
since the parallel layers only occupy a region of width $a$ (equal to several lamellar
widths). It is easy to show that $C_\parallel<C_m<C_\perp$. 

The free energies per unit area of the three competing phases are given by
\begin{eqnarray}
F^\parallel &=&Lf_p+2\gamma_{_{\rm AS}}-\frac12 C_\parallel L^2E_0^2~,\\
F^\perp &=&Lf_p+\gamma_{_{\rm AS}}+\gamma_{_{\rm BS}}-\frac12 C_\perp L^2E_0^2~,\\
F^m &=&Lf_p+2\gamma_{_{\rm AS}}+2\gamma_{_T}-\frac12 C_m L^2E_0^2~.
\end{eqnarray} 
$f_p$ is the polymer free energy per unit volume and includes bending and
compression.
$\gamma_{_{\rm AS}}$ and $\gamma_{_{\rm BS}}$  are the A and B interfacial
energies with the substrate.
\begin{figure}[h!]
\includegraphics[scale=0.52,bb=45 577 540 775,clip]{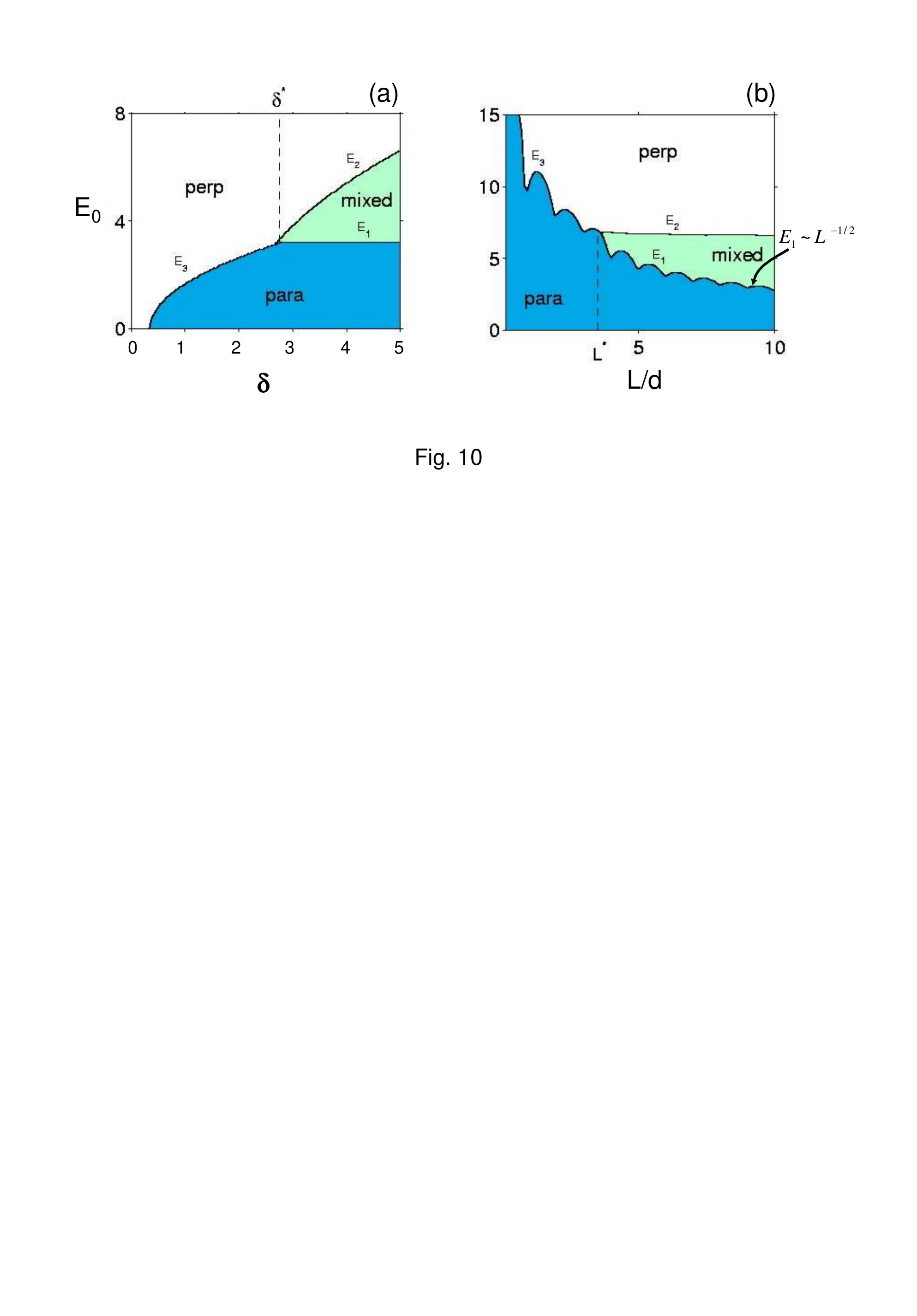}
\caption{Stability diagrams for confined lamellar phases in the strong
segregation regime
in electric fields. (a) $L=10d$ and is fixed. Small value
of $E_0$ leads to parallel layers. Increase in $E_0$ leads to a direct
transition to
perpendicular layers when $\delta=(\gamma_{_{\rm
BS}}-\gamma_{_{\rm AS}})/\gamma_{_T}$ is small enough. If
$\delta$ is large, $\delta>\delta^*$, increase in the field induces first a
transition to
a mixed phase and subsequently to a perpendicular orientation. (b) Diagram in
the $E_0$-$L$ plane with fixed $\delta=5$. $E_0$ is scaled by
$(\gamma_{_T}/d)^{1/2}$.
Adapted with permission from \cite{TA_mm2002}. Copyright
2002 American Chemical society.
}
\label{fig_orient_diag}
\end{figure}

The minimal model above allows the calculation of phase diagrams as a function of $E_0$, 
$\gamma_{_{\rm BS}}-\gamma_{_{\rm AS}}$, and $L$. Figure~\ref{fig_orient_diag} shows two
such cuts in the three-dimensional phase space. The transition fields $E_1$, $E_2$ and
$E_3$ are lines where the free energies are equal to each other. In part (b), the
transition field $E_1$, above which parallel lamellae become
unstable, has undulations. These are due to the incommensurability between the lamellar
period and the film thickness, causing polymer stretching or compression.
$E_1$ decays like $L^{-1/2}$ ~-- this is easily understood since the interfacial
energy is
proportional to the area, while the electrostatic energy
scales as area $\times LE_0^2$. $E_2$ is the transition field between mixed and
perpendicular
lamellae and is virtually independent
of film thickness $L$. Note that $|\gamma_{_{BS}}-\gamma_{_{AS}}|$ may be so
large that
$E_2$ is above the threshold for dielectric breakdown ($\sim 150$ V/$\mu$m). In
these circumstances perfect perpendicular lamellae throughout the whole film are
not possible.

\begin{figure}[h!]
\includegraphics[scale=0.7,bb=122 625 450 785,clip]{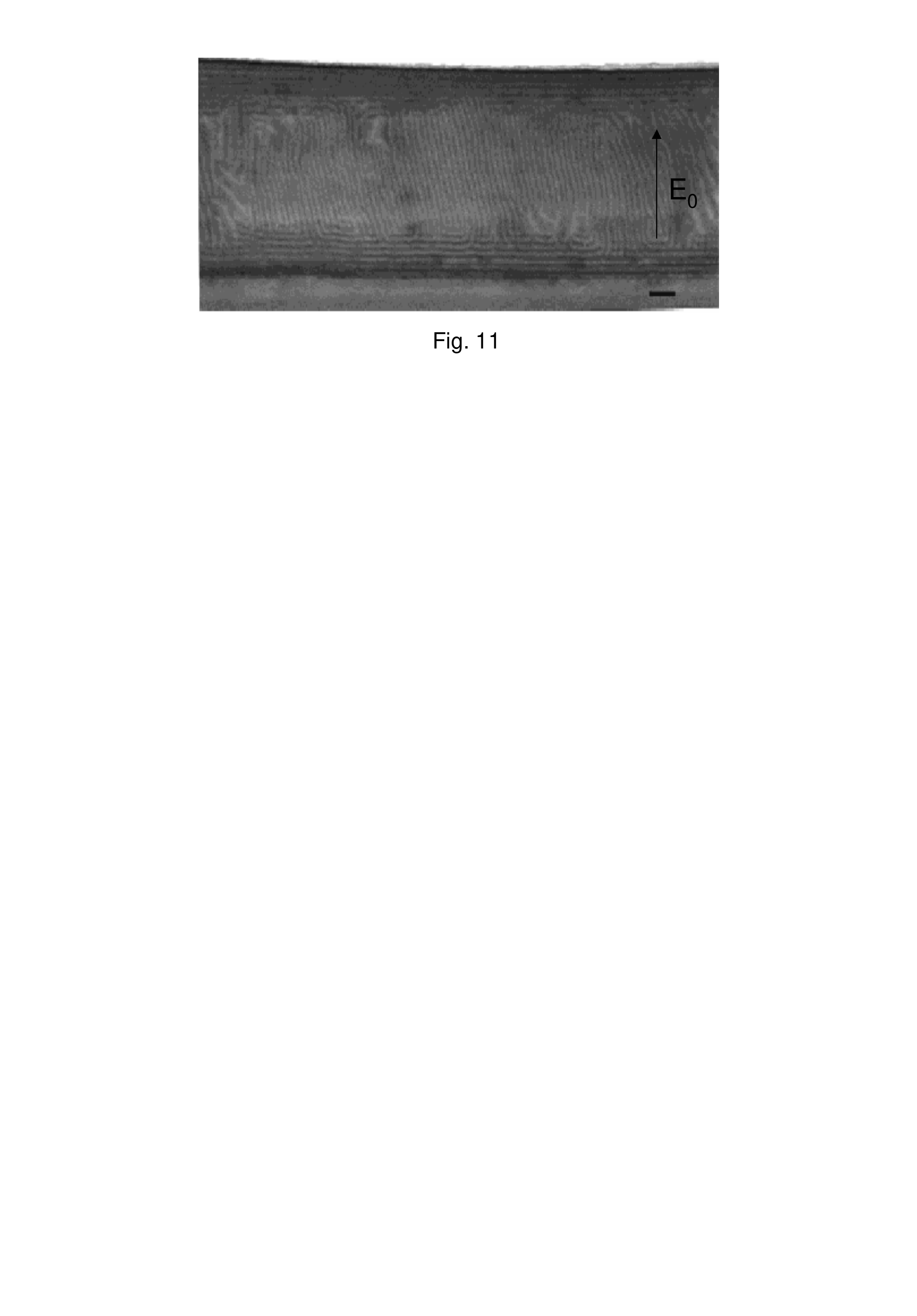}
\caption{TEM cross section of thin-film PS/PMMA block copolymer annealed in
electric
field $E_0\simeq 40$ V/$\mu$m. Few lamellae lie parallel to the substrate
(bottom) and
polymer-air interface (top), while the rest of the film has lamellae parallel to the
field.
Adapted with permission from \cite{russell_mm2004}. Copyright
2002 American Chemical society.
}
\label{fig_Tjunction}
\end{figure}

Experiments validate the existence of a mixed morphology. Fig.
\ref{fig_Tjunction} shows lamellae sandwiched in a thin film in the presence of
a perpendicular electric field. Clearly, the film's center has lamellae oriented
parallel to the field's direction, while close to the interface with the
substrate or air, the copolymer component with lowest interfacial energy
is preferentially adsorbed, leading to the creation of few layers parallel to
the substrate.
The defect separating the two regions, while not perfectly similar to the ideal
one depicted in Fig.~\ref{fig_para_perp_mix}, can nonetheless be associated with
an energy $\gamma_{_T}$ per unit area so the model should still be valid.

Muthukumar has considered other confined ordered phases in
electric field by using a similar model. While the dependence of the
transition field between parallel and perpendicular orientations on $L$ is weak for
lamellae, they find that for cylinders the $L$ dependence of the energy is more
complicated, and the transition field changes markedly as the film thickness changes
\cite{muthu_jcp2001}.

\subsubsection{Electrostatics of weakly segregated structures}

The above derivations are valid in materials (not necessarily block copolymers)
where the dielectric interfaces are sharp and $\Delta\eps$ is large compared to
the average 
permittivity $\bar{\eps}$. Close to the
critical point, however, the composition differences between coexisting phases
become small. The expressions for the electric field and electrostatic 
energy given below
only assume small $\eps({\bf r})$ variations and therefore are general and not 
restricted to BCPs.
The smallness of $\eps$ variations, as compared to $\bar{\eps}$, 
enables us to exactly solve Laplace's equation for an {\it arbitrary} spatial
distribution $\eps({\bf r})$.
Close to the critical point, we may write the order parameter as a sum of Fourier
harmonics [Eq. (\ref{op_fourier})],
with $\vphi({\bf r})\ll 1$. Consequently, only linear terms are retained 
in the constitutive relation $\eps(\phi)$ and in the expansion of ${\bf E}$ in powers
of $\vphi$ [Eqs. (\ref{const_relation}) and (\ref{E_expansion})].
The deviation of the electric field from its
average, ${\bf E}_1\vphi$ in Eq. (\ref{E_expansion}), is found by writing Laplace's
equation $\nabla\cdot(\eps{\bf E})=0$ to linear order in $\vphi$ and is given
by
\cite{ah_mm1993},
\begin{eqnarray}\label{deviation_field_E1}
\vphi{\bf E}_1&=&\sum_{\bf q}{\bf E_q}e^{i{\bf q}\cdot{\bf r}}~,\\
{\bf E}_{\bf q}&=&-\frac{\Delta\eps\vphi_{\bf q}}{\bar{\eps}}\left(\hat{\bf
q}\cdot{\bf
E}_0\right)\hat{\bf q}~.\nn
\end{eqnarray} 
The electrostatic energy per unit volume can be approximated by
\begin{equation}\label{ah_fe_expression}
f_{\rm es}=\frac{(\Delta\eps)^2}{2\bar{\eps}}\sum_{\bf q}(\hat{\bf q}\cdot{\bf
E}_0)^2\varphi_{\bf q}\varphi_{-{\bf q}}+{\rm const.}
\end{equation} 
The constant on the right-hand side is the electrostatic energy for a uniform
phase with
$\phi({\bf r})=\phi_0$. Any plane-wave composition deviation $\varphi_{\bf
q}e^{i{\bf q}\cdot{\bf r}}$ from the isotropic
background leads to an electrostatic penalty. This energy, quadratic in $E_0$ as
usual, 
is also quadratic in $\vphi_{\bf q}$. This is because in uniform electric
fields, composition fluctuations
$\vphi$ give rise to dielectric deviations $\delta\eps\propto\vphi$ and
electric-field deviations $\delta E\propto\vphi$, so the leading order term in the
free energy density $-(1/2)\eps E^2$ scales as $\vphi^2$.

The torque ${\bf N}$ acting on a sample of volume $\cal{V}$ in external field
${\bf E}_0$ consistent
with the energy formula [Eq. (\ref{ah_fe_expression})] is \cite{tsori_mm2003}:
\begin{eqnarray}\label{torque_dielec}
{\bf N}_{\rm dielec}&=&-2\mathcal{V}\frac{(\Delta\eps)^2}{\bar{\eps}}
{\bf \Gamma}(\phi,{\bf E}_0)\\
{\bf \Gamma}(\phi,{\bf E}_0)&=&
\sum_{\bf q} \varphi_{\bf q}\varphi_{-{\bf q}}
(\hat{\bf q}\cdot{\bf
E}_0)\hat{\bf q}\times {\bf E}_0~.\nn
\end{eqnarray} 
An important feature of Eqs. (\ref{ah_fe_expression}) and (\ref{torque_dielec}) is
 the dependence on the {\it angle} between ${\bf q}$ and ${\bf E}_0$. If ${\bf q}$ and
${\bf
E}_0$ are perpendicular to each other (dielectric interfaces parallel to ${\bf E}_0$), the
energy is minimal and the torque vanishes. When ${\bf q}$ and ${\bf E}_0$ are parallel 
(interfaces perpendicular to ${\bf E}_0$), the torque
vanishes, but the energy is highest.

Equation (\ref{ah_fe_expression}) is a useful formula for the calculation of
electrostatic
effects in soft-matter systems close to a critical point. When it complements
a proper Ginzburg-Landau expansion of the energy in powers of $\vphi$, it
enables
rather easy derivation of various thermodynamical expressions on the mean-field level
\cite{onuki_book}.
This issue will be discussed in Sec. \ref{sect_crit_eff}.

\subsubsection{Numerical calculations}

A numerical approach for the calculation of electrostatic effects in spatially nonuniform 
materials has the advantage that it does not assume that $\Delta\eps/\bar{\eps}$ 
is small or large.
Numerical calculations therefore bridge the
gap between weak- and strong-segregation theories outlined above. Electrostatic
calculations 
for block copolymers have been so far implemented by mean-field self-consistent field
theory (SCFT).
In SCFT of block copolymers, the Hamiltonian contains chain stretching penalty and
enthalpic
interactions, originating from unfavorable contacts between chemically different monomers.
The partition function is expressed in terms of the density 
operators of the different types of monomers. The so-called restricted chain partition
function 
obeys a modified diffusion equation \cite{matsen_schick_prl1994}. In order to
account for electrostatic effects, a 
constitutive relation $\eps(\phi)$ is chosen (usually a linear relation), and 
the electrostatic energy
$F_{\rm es}$ in Eq. (\ref{f_es}) is added to the Hamiltonian.
The full set of equations must be solved self-consistently together with Laplace's
equation and subjected to the proper boundary conditions on the 
electrodes, i.e., Dirichlet, Neumann, or mixed boundary conditions 
\cite{schick_mm2005,matsen_mm2006}.
The main advantage of the formalism for this study is the exact calculation of
electrostatic energy - the numerical procedure does not rely on the analytical
approximations
such as Eq. (\ref{ah_fe_expression}).

While static calculation pertains to equilibrium morphologies, a certain variant of SCFT
allows description of the dynamical relaxation toward equilibrium. In the scheme
developed 
\cite{fraaije_jchemphys1999}, the free energy is calculated similarly to the
static case. 
The time variation in the
order parameter $\phi$ is then assumed to obey the following dynamics \cite{onuki_book}:
\begin{eqnarray}
\frac{\partial\phi}{\partial t}=M\nabla^2\mu+\eta~,
\end{eqnarray}
where the chemical potential $\mu$ is a functional derivative of the total free energy 
including $F_{\rm es}$, $M$ 
is a proper Onsager mobility coefficient (usually taken to be constant), and $\eta$ is a
noise
term satisfying the fluctuation-dissipation theorem.
This approach does not fully take into account the polymer viscoelastic flow
and reptation effects. In addition, the Amundson-Helfand quadratic
electrostatic energy approximation [Eq. (\ref{ah_fe_expression})] was
used instead of the full expression \cite{boker_soft_matter2007}. 
Still, a reasonable match with experiment was
found: the simulations report an exponential relaxation of
the order parameter $S(t)$ [Eq. (\ref{S_op})], consistent with experiments
\cite{boker_mm2003}.

It should be stressed that the approach utilized above, that $\eps$ is
a simple scalar quantity, may fail with some polymers. Recent experiments show a
peculiar phenomena, whereby the lamellar period changes by as much as $5\%$ 
under the influence of an electric field {\it parallel}
to the lamellae \cite{boker_nat_mat2008}. The researchers attribute the thinning
of lamellae to an anisotropic response of the polymers (PS/PI). In essence, the
single chain
conformations of the PI block are affected by the field, and the coil becomes
more prolate
in a direction parallel to the field \cite{guro_mm1994}.

Lastly, the cylinders in Fig.~\ref{fig_bcp_orient}(c) follow the {\it
curved} electric field lines. The electric field introduces electrostatic
(Maxwell)
stress in the anisotropic sample, and this stress must be balanced by elastic
forces. The competition between electrostatic and elastic forces has not
received enough attention in the literature and should be further investigated.

\subsection{Instability of block-copolymer interface in perpendicular electric
field}

In light of the normal-field instabilities described in Sec.
\ref{interf_instab}, 
one may ask whether a similar interfacial instability can also occur in
block copolymers.
The difference between self-assembled materials, such as block copolymers, and
simple liquids is that besides
their high viscosity they feature {\it elastic} behavior. Indeed lamellar phases, for
example,
have both bending and compression moduli. As a result, any deformation of the interface
between the
copolymers results in deviation from the optimum balance between interfacial and entropic
energies.
It turns out that an interfacial instability does occur in BCPs in electric field, 
and the nature
of the instability depends on the distance from the critical point.

\begin{figure}[h!]
\includegraphics[scale=0.55,bb=50 645 515 790,clip]{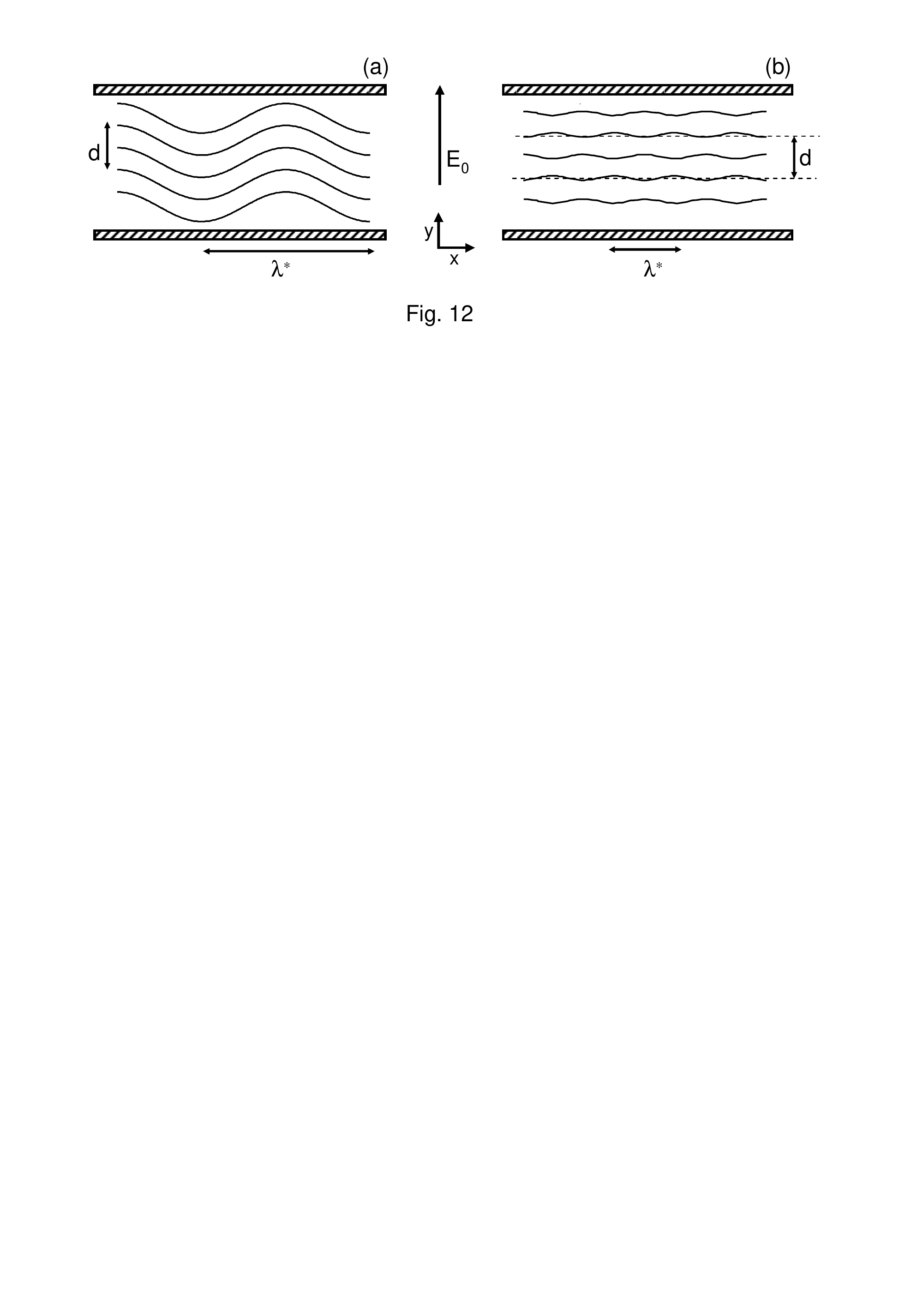}
\caption{Schematic illustration of two normal-field instabilities in lamellar block
copolymers
(not to scale).
$d$ is the unperturbed lamellar thickness and $\lambda^*$ is the undulation period.
(a) In the strong-segregation regime, adjacent lamellae bend in phase with each other.
$\lambda^*$ is macroscopic, $\lambda^*\gg d$.
(b) Close to the ODT point, lamellae are easier to deform, and adjacent lamellae undulate 
out of phase with each other.
$\lambda^*$ is equal to the lamellar thickness $d$.
}
\label{fig_bcp_instab}
\end{figure}

\textcite{onuki_mm1995} studied the strong-segregation regime by writing 
the copolymer composition as $\phi(x,z)=\phi_0+\vphi_L\cos[(2\pi/d)x+u(y)]$, 
with $\vphi_L$ as an
amplitude and $d$ as the lamellar period,
and expanding the energy
in small $u$. They found an interfacial instability with two characteristics: (i)
two adjacent A/B polymer interfaces are parallel to each other and (ii) the most unstable
wavelength is long, corresponding to the lateral film size [Fig.
\ref{fig_bcp_instab}(a)]. 
This instability is similar to
the Helfrich-Hurault instability common in smectic and cholesteric liquid crystals 
\cite{pgg_prost_book}.

A different scenario occurs close to the ODT (critical) point. In this regime, the
composition
variations are very weak, and lamellae are easy to deform. A stability analysis of the
Ginzburg-Landau-like free energy expansion, appropriate close to a critical
point and on the mean-field level,
reveals that the most unstable wavelength $\lambda^*$ is the same as the bulk
period $d$
[see Fig.~\ref{fig_bcp_instab}(b)] \cite{TA_mm2002}. 
This is very different from the instability in the liquid case.
Moreover, adjacent lamellae are out of phase with each
other. At the onset of instability, the BCP morphology is essentially a superposition of
two lamellar phases with normals parallel and perpendicular to the substrate.
These modulations were also corroborated by a more accurate SCFT study 
\cite{matsen_prl2005,matsen_softmatter2006}, which describes undulatory ``peristaltic''
modes
similar to that of Fig.~\ref{fig_bcp_instab}.
These theoretical predictions were essentially verified in experiment
\cite{russell_cps2004}.

\subsection{Role of residual ions}\label{mobile_ions}

The preceding discussion of block copolymers assumes that the polymers are purely
dielectric materials and ignores the possible existence of dissociated ions.
However, almost all polymers have small conductivity due to the finite amount of 
ions in them. For example, in anionic polymerization the reaction typically starts 
with an organometallic reagent, such as butyl lithium. At the end of the reaction, 
each chain is neutralized with water, and one metal hydroxide ion (e.g., LiOH) is 
released. Thus, the average ion density is one ion per polymer chain, which is equivalent
to
a number density of $10^{25}$ m$^{-3}$ ($\simeq 0.016$M) or charge
density of $1.6\times 10^6$ C m$^{-3}$.

Even without electric fields,
a salt content adds to the molecular polarizability of the polymer, and therefore changes
the interaction parameter $\chi$ between polymer molecules. In addition, like any
chemical impurity, salt can change the interfacial interactions of the polymers with
the substrate.
The effect of ions on the BCP
phase diagram was studied quantitatively
\cite{bates_chemmater2002,bates_mm2003,russell_mm2008}. For a given
phase, we are interested in the effect ions have on the orientation process in external
fields.
Most of the salt content crystallizes and is found in neutral clusters inside the
polymer, and only a small fraction ($10^{-3}-10^{-6}$) is dissociated and 
participates in the orientation
mechanism. Of those ions, some are complexed to the polymer to which they have a
preferential solubility, usually the more polar block. They increase the effective
dielectric constant of the polymer and therefore increase the effective
$\Delta\eps$. 

The
rest of the ions are mobile; 
when subjected to electric field, they give rise to screening. 
Two extreme cases follow:
in the first, the screening length $\lambda$ is much larger than the lamellar period
$d$ and in fact may even be as large as the film
thickness. When this happens, the ion cloud is uniformly distributed throughout
the film in the more polar polymer, and
the system behavior is similar to that of neat BCPs, only with effective
$\eps$'s.
The critical field for transition from parallel to perpendicular lamellae
still scales as $E\sim L^{-1/2}$.
Orientation experiments
were carried out on
PS/PMMA samples doped with ions 
\cite{russell_cps2004,russell_prl2006}. Ions reside preferentially in the 
PMMA block,
resulting in an increased $\Delta\eps$ and correspondingly
smaller fields for the orientation process.
How much the dielectric constant of the PMMA block is enhanced depends on the 
frequency of applied field and on the temperature \cite{ttalbrecht_prl2009}. 
Typically, the dielectric
constant increases with decreasing frequency. Large doping with ions can lead to 
even a fivefold increase in the static dielectric constant.

The second extreme case corresponds to screening length $\lambda$ smaller than $d$ 
and occurs when salt is added. In
this case, most of the ions concentrate at the electrodes. Here one expects 
the most polar
polymer to wet the surface due to the preferential solubility of ions. The effect of
electric field is thus reminiscent of electrowetting and can be approximately 
described
by an effective interfacial energy between the polar block and the substrate. 
Note that
in contrast to the case with neat BCPs, increase in the electric field increases the
tendency of lamellae wetting
parallel to the substrate. The effective surface tension depends on the total 
ion density
and vanishes with vanishing electrode potential $V$.
The mixed structure depicted in Fig.~\ref{fig_para_perp_mix} is
possible for intermediate screening lengths $\lambda\geq d$ and depends on the
vicinity to the ODT point.

The physics of the problem is rich when the external field varies sinusoidally in time.
When a time-varying electric field is applied, the positive charge will tend to migrate
parallel to the field, while negative charge will move in the opposite direction in an
oscillatory motion.
There are four length scales in the problem: the BCP
domain size is $d\sim50$-$100$ nm, comparable to the polymer radius of gyration.
The second is the electromagnetic wavelength, 
$2\pi c/(\omega\sqrt{\bar{\eps}})$,
where $\omega$ is the field's frequency.
The third length is the distance a single ion drifts in one half period
of the electric field: $\pi e\bar{\mu} E_0/\omega$, where $\bar{\mu}$ is 
the average mobility. 
Lithium mobility in PMMA is estimated to be $\bar{\mu}\simeq (4$--$5)\times 10^5$ m$^2$/J
s.
Typically the wavelength of light is much larger than both 
the drift
length and the BCP period. Lastly, the system size $L$ is usually the largest length. 
There are three energy scales: one is the energy stored in the dielectric material 
per unit volume, $U_{\rm dielec}=\bar{\eps}E_0^2$. The second energy is the thermal
energy $k_BT$. The third energy is the amount
of heat dissipated per unit volume due to Joule heating in one field cycle: 
$U_{\rm Joule}=2\pi E_0^2\sigma/\omega$. Using an average density of dissociated
Li ions $n \sim 10^{21}$ m$^{-3}$, one obtains 
the average conductivity $\sigma=e^2n\bar{\mu}\sim 10^{-11}$ C/m V s.
Thus, Joule heating begins to be important at low
frequencies, $\omega <\omega_c$ [Eq. (\ref{omega_c})], where here
$\omega_c\sim 5$--$50$ s$^{-1}$.

The copolymers are anisotropic materials, and therefore the mobility of ions is
spatially nonuniform. Ions will consequently drift in the direction which is ``easy''
for them, and this direction may not be collinear with the external field. 
As a result, a {\it torque} is applied on the sample. The total torque 
works to orient the sample in such a direction that it vanishes and 
the mechanical energy is at minimum.
Clearly, the torque depends on the external field's frequency: small $\omega$
and long period means ions have a long way to drift before the field changes sign,
and hence the torque exerted on the polymer matrix is large. 
At large values of $\omega$, ions are practically immobile and contribute nothing
to orientation. A detailed calculation shows that the torque exerted by moving ions
is dominant over the ``dielectric interfaces'' if $\sigma/\omega\gg \bar{\eps}$.
An expressions for the torque in this regime exists, and it reads as \cite{tsori_mm2003}
\begin{eqnarray}\label{torque_ions}
{\bf N}_{\rm ions}=
-\frac{8\sigma}{\pi\omega}\mathcal{V}\left(\frac{\Delta\mu}{\bar{\mu}}\right)^2
{\bf \Gamma}(\phi,{\bf E}_0)~.
\end{eqnarray} 

This formula should be contrasted with Eq. (\ref{torque_dielec}). 
It is valid close to the critical point and under the assumptions that positive and 
negative charges have equal mobilities and using a linear dependence of mobility 
on composition: $\mu(\phi)=\bar{\mu}+\varphi\Delta\mu$.
In this regime, occurring when $\omega<\omega_c$, 
the mobile ions become increasingly important and the concept of dielectric
contrast
should be replaced by mobility contrast.
In this regime the sample is being heated; in most experimental systems,
however, this energy dissipation does not present a problem since heat is quickly
removed from the system.

The stresses introduced by moving ions and by dielectric interfaces 
(Maxwell stress) can lead to real symmetry-changing phase transitions and not
just to orientation. This will be discussed in subsequent sections.

\section{Critical Effects in Polymer and Liquid Mixtures in Uniform Electric
Fields}
\label{sect_crit_eff}

Up to this point, we described several instabilities and orientation effects
occurring when two immiscible
liquids or ordered phases of block copolymers are subjected to an external electric field.
One may ask even a more basic 
question: how does an electric field affect the relative thermodynamic stability of the
possible system phases? We turn to investigate this question on the mean-field level as
is the
practice throughout this text.

\subsection{Landau theory and experiments}

We consider a bistable system, able of having two different states at a certain range
of the parameters. For concreteness one may think of a binary mixture of two liquids.
The treatment of
liquid-vapor coexistence of pure molecules is quite similar, the main difference being
the finite gas compressibility which means that the Clausius-Mossotti relation connects
between $\eps$ and density \cite{moldover_prl1999,moldover_pre1999,amara_prl2004}.
What is the effect of a uniform electric 
field on the phase diagram of liquid mixtures?
This issue was treated first by \textcite{LL_book_elec}. 
For simplicity of exposition, we consider a symmetric liquid mixture. 
The free energy
density in the absence of field, $f_m$, can be written as a Landau expansion valid
close to the critical point,
\begin{eqnarray}\label{fm}
\frac{v_0}{k_BT_c}f_m=\frac12\frac{T-T_c}{T_c}\varphi^2+\frac{u}{4!}\varphi^4~.
\end{eqnarray}
Here $\varphi$ is the deviation of the composition from the critical composition ($1/2$),
 $v_0$ is a volume derived from the second virial coefficient, and $u$ is
positive.
We use the constitutive relation [Eq. (\ref{const_relation})] to write
the electrostatic energy density 
$f_{\rm es}\simeq -\frac12
 \eps_c
E_0^2-\frac12\Delta\eps\varphi E_0^2-\frac14\eps''\varphi^2 E_0^2$.
The constant term is unimportant and the linear term can be eliminated by a 
redefinition of the chemical potential. The only important term is the quadratic one.
Inspection of $f=f_m+f_{\rm es}$ shows that since $f_{\rm es}$ is small, close enough to
$T_c$,
that $f_{\rm es}$ simply serves to redefine
the critical temperature. Thus
\begin{eqnarray}
T_c\to T_c+\Delta T_c~,\nn\\
\Delta T_c=\frac{v_0\eps''E_0^2}{2k_B}~.
\label{DT_landau_mechanism}
\end{eqnarray}
$T_c$ and the whole binodal curve are increased if $\eps''$ is positive, and in
this case electric field leads
to {\it demixing}. If $\eps''<0$, then $T_c$ is decreased and {\it mixing} is
favored.
We use the value $v_0\sim 10^{-28}$ m$^3$ and average electric field $E_0\sim10$
V/$\mu$m
to obtain the typical shift to
$T_c$ due uniform fields in either case to be $v_0\eps'' E_0^2/k_B\sim 30$ mK.
Subsequent theoretical work employed a renormalization-group \cite{onuki_epl1995} 
and other approaches \cite{sbmg_physicaa1980} 
to study the vicinity of the critical point, but the
main result remained: $T_c$ changes by few milikelvins under the influence of
moderate-to-strong fields.

The theory should be compared with experiments which followed. The first experiment
was by
\textcite{debye_jcp1965} on mixture of isooctane and nitrobenzene, whose dielectric
contrast is $\Delta\eps=32.3\eps_0$. Under an electric field of $4.5$ V/$\mu$m, they
found that $T_c$ is reduced by $15$ mK. These results were later verified with
great accuracy by \textcite{orzech_chemphys1999}. Debye's experiments were followed by a
work of \textcite{beaglehole_jcp1981}. He worked on a cyclohexane/aniline mixture with
$\Delta\eps=5.8$ and in a field of $0.3$ V/$\mu$m. $T_c$ was found to be reduced by as
much as $80$ mK. \textcite{early_jcp1992} worked on the same mixture with similar fields.
He found that $T_c$ does not change at all and hinted that previous results were due to
spurious heating occurring in dc fields. \textcite{wirtz_fuller_prl1993} worked on
n-hexane/nitroethane mixture with $\Delta\eps=17.7\eps_0$ and found a reduction in $T_c$
by $20$ mK.

The experiments above are in contradiction to theory since mixing is observed instead
of demixing even though $\eps''>0$. 
In addition, there were wrong signs in some experiments as well as theory. More
importantly, in both theory and experiment the change to $T_c$ is much smaller than $100$
mK. The only exception is the work of \textcite{reich_jpspp1979}, who worked on a
polymer-polymer mixture having a lower critical solution temperature (LCST) and found that
$T_c$ changes by about $4$ K or even more under an electric field of $\sim 10$ V/$\mu$m.
Two comments on the big effect observed are the following: (i) The large molecular volume
of polymers
means their entropy is reduced. $v_0$ should be replaced by $Nv_0$ in Eq.
(\ref{DT_landau_mechanism}). 
(ii) Heating, if present
in a LCST system in the homogeneous phase, may lead to demixing, as is observed
in these experiments.

A possible explanation for the mixing observed in experiments is due to the dielectric
anisotropy not accounted for by Eq. (\ref{DT_landau_mechanism}). 
Indeed, just as with block copolymers or any other anisotropic
material, composition variations $\vphi$ lead to dielectric constant variations
$\delta\eps$. Compliance to Laplace's equations means that the electric field has
variations
too, and in linear order $\delta E\propto -\delta\eps$. As is clear from Eqs.
(\ref{op_fourier}) and (\ref{ah_fe_expression}), the
electrostatic free energy contribution $f_{\rm es}$ scales quadratically with $\varphi$.
We
therefore conclude that the prefactor of $\vphi^2$ in a Landau series expansion
of
$f_m+f_{\rm es}$ around the critical composition has two terms: one proportional to
$-\eps''$ and the second to $+(\Delta\eps)^2$. A positive $\eps''$ favors demixing while
$(\Delta\eps)^2$ term promotes mixing and is usually dominant.

\vspace{0.5cm}
\noindent
{\bf Light-induced phase transitions in mixtures}\\

We briefly mention several cases where light influences the morphology of mixtures.
Light in the UV range was applied on polymer blends of stilbene-labeled polystyrene
and polyvinylmethylether (PSS/PVME).
This caused phase separation of the blend because {\it trans}$\to${\it cis} transition
changes both the volume and the enthalpic interaction between the polymers
\cite{qui_mm2000}.  
In addition, 
photochemical reaction was induced in anthracene-labeled polystyrene-PVME blends
irradiated by periodic UV light. The distribution of length scales in the resulting
spinodal decomposition depends sensitively on the applied frequency
\cite{qui_nature_mat2004}.

Even in the absence of chemical activity, an intense laser beam can lead to a mixing or
 demixing
phase transition. Poly($N$-isopropylacrylamide) (PNIPAM) --
water and
PNIPAM -- D$_2$O mixtures were illuminated by tightly focused laser light in the infrared
wavelength \cite{masuhara_bcsjapan1996,masuhara_langmuir1997}. These mixtures have LCST
(inverted) phase diagrams and the initial temperature corresponded to a homogeneous state.
Aggregation of the polymer
was observed in the region illuminated by the laser. Two possible explanations were
suggested. First, water adsorption in the infrared regime causes heating, and in a LCST
system this may cause phase separation. Second, radiation pressure supposedly works
in conjunction with the heating to separate the polymer from the solvent and to
accumulate it in micron-scale aggregates. Subsequent work
\cite{masuhara_jphyschemB1997} was similar,
but now poly($N$-vinylcarbazole) was dissolved in two different organic solvents,
cyclohexanone and $N$,$N$-dimethylformamide.
This work showed that the
segregation of polymer from the solvent occurs in the absence of heating and is due to
the strong laser electric field.
The underlying physical mechanisms responsible for the observations were
theoretically examined by \textcite{ducasse_physicaa1999}. In Sec. 
\ref{sect_efips} we give a possible mechanism which may be relevant to equilibrium
situations.

\subsection{Block-copolymer phase transitions}

In Sec. \ref{sect_bcp}, block copolymer orientation in electric fields was described as
a
result of a ``dielectric interfaces'', stemming from a term proportional to
$(\Delta\eps)^2$
in the free energy. Equation
(\ref{torque_dielec}) shows that a BCP sample experiences a force as long as there are 
dielectric interfaces perpendicular to the external field. Phases with uniaxial symmetry
thus turn until their axis of symmetry is parallel to ${\bf E}_0$. When this
orientation terminates, their
total electrostatic energy is exactly equal to $-\frac12\bar{\eps}E_0^2\times
\mathcal{V}$,
where $\mathcal{V}$ is 
the total volume. But there are ordered phases, such as the bcc phase of spheres, where 
frustration always occurs. As $E_0$ is increased, the crystal deforms, and the spheres
elongate along the external field. This is reminiscent of the liquid droplet elongation
described in Sec. \ref{droplet_elongation}, but in addition to the interfacial tension
and electrostatic energies, in BCPs there is also elastic energy that needs to be taken
into
account.
With increasing value of $E_0$, the crystal deforms, until, at a certain critical value
of 
$E_0$, its energy is no longer the lowest among all phases, and a phase transition occurs
\cite{tsori_prl2003}.

The phase diagram close to the critical point can be calculated by a Landau expansion of
the
energy in series
of $\varphi_{\bf q}$, given in Fourier space by
\begin{eqnarray}\label{f_es_expan_bcp}
\frac{v_0}{k_BT_c}F&=&\frac12\sum_{\bf q}
\left[\frac{T-T_c}{T_c}+\frac{v_0(\Delta\eps)^2}{\bar{\eps}k_BT_c}
(\hat{\bf q}\cdot{\bf E}_0)^2\right.\nn\\
&+&\left.C(q^2-q_0^2)^2\right]\varphi_{\bf q}\varphi_{-{\bf q}}+O(\varphi_{\bf q}^3)~.~~~
\end{eqnarray}
The last term in the square brackets is specific to modulated phases
such as block copolymers, and is absent in simple liquids. It expresses
the penalty of having $q$ modes with $q$ different than the preferred inverse periodicity,
$q_0=2\pi/d$. The prefactor $C$ contains information on the polymer, such as the radius of
gyration.
The electric field (second term) introduces a penalty proportional to the cosine
squared of the angle between
field and $q$ vector. As is mentioned above, this field-dependent term is usually quite
small:
when we replace $\eps''$ by $(\Delta\eps)^2/\bar{\eps}$ in the estimate after 
Eq. (\ref{DT_landau_mechanism}), we get
the same millikelvin-sized effect of renormalization of $T_c$.
For block copolymers, however, the molecular weight can be large and thus, replacing $v_0$
by
$Nv_0$, one may get a larger change to $T_c$, similar to the experiments on polymer
mixtures \cite{reich_jpspp1979}.

\begin{figure}[h!]
\includegraphics[scale=0.45,bb=25 555 550 770,clip]{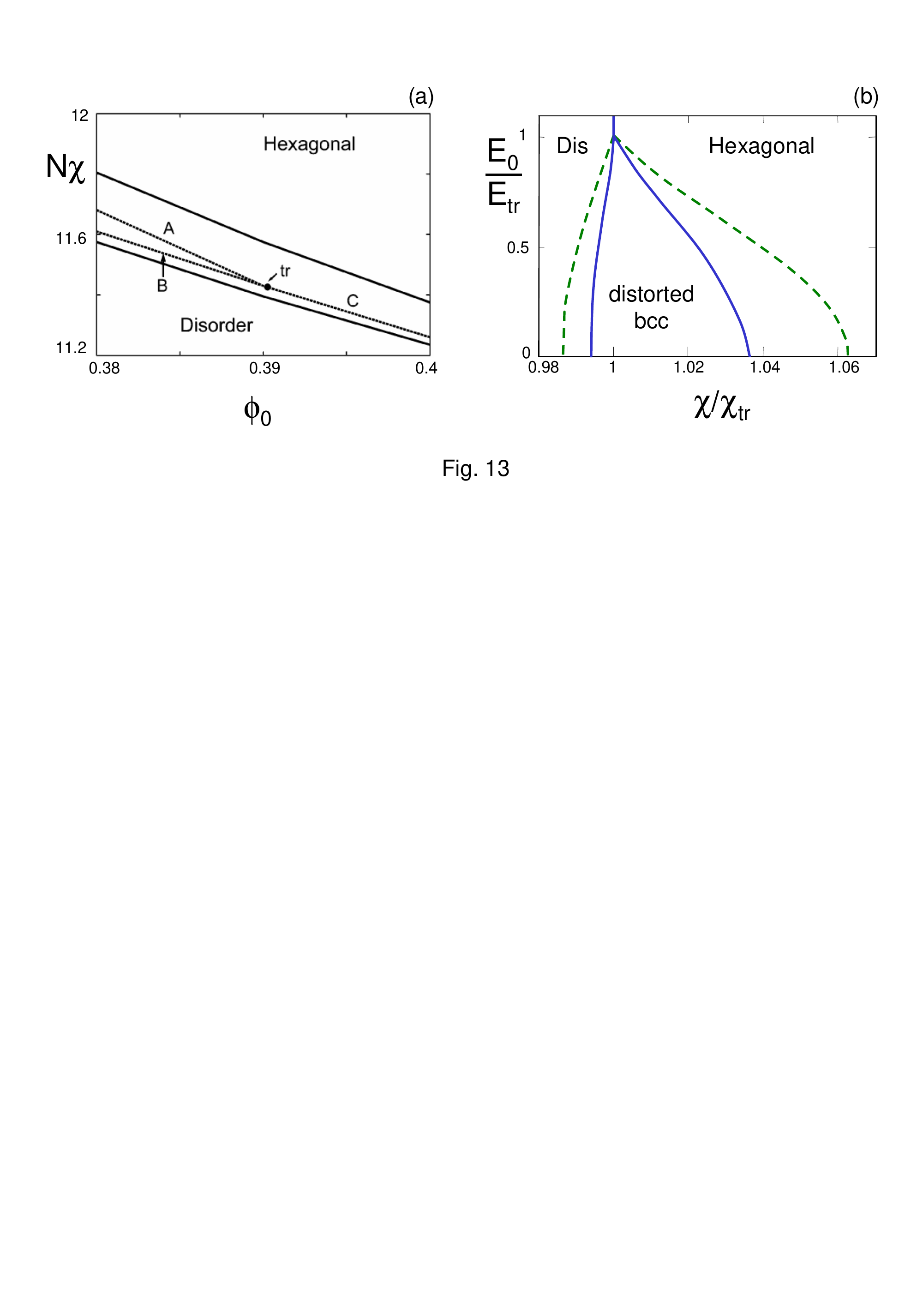}
\caption{Block-copolymer phase diagrams. 
(a) Phase diagram for diblock copolymers in the ($\phi_0$,$\chi$) plane
at
fixed electric field. Two outer solid lines are zero-field hexagonal-to-bcc and
bcc-to-disorder phase boundaries. 
Lines A and B are transitions from distorted bcc to the 
hexagonal and disordered phases, respectively, and line C is the boundary between
hexagonal and disordered phases in electric field given by
$(\eps_0Nv_0/k_BT)^{1/2}E_0=0.2$. 
They meet at the triple
point ($\phi_0$,$N\chi$)=($0.39$,$11.43$). Adapted from \cite{schick_mm2005}.
(b) Phase diagram in the ($E_0$,$\chi$) plane and fixed composition $\phi_0=0.3$.
For low values of field, a distorted bcc phase lies
between the disordered (``Dis'') and hexagonal phases. The phase lines
terminate at a triple point ($\chi_{\rm tr}$, $E_{\rm tr}$). When $E_0>E_{\rm tr}$, the
distorted bcc does not exist. Solid lines are from analytical one-mode approximation, 
whereas dashed lines are from self-consistent field theory. The axes of each
calculation 
are scaled by the relevant values of $\chi_{\rm tr}$ and $E_{\rm tr}$. 
Adapted with permission from \cite{schick_mm2005} and \cite{tsori_mm2006}.
Copyright 2002 American Chemical society.
}
\label{fig_bcp_diag_field}
\end{figure}
Figure~\ref{fig_bcp_diag_field}(a) is a phase diagram in the ($\phi_0$,$\chi$) plane
($\chi\sim 1/T$) for fixed electric field, obtained with the SCFT approximation. A
linear constitutive relation was used:
$\eps(\phi)=\bar{\eps}+\Delta\eps\varphi$.
Solid
lines are zero-field phase boundaries, while dotted lines marked with A, B, and C are the
new phase boundaries in the presence of an external field. There appears a triple point
where three phases meet. At values of $\phi_0$ smaller than the triple point value, a
direct transition between hexagonal and disordered phases occurs as a function of $\chi$
(or $T$).
Part (b) is a phase diagram in the ($\chi$, $E_0$) plane
for fixed $\phi_0$. 
Dashed lines are a result of SCFT calculation, while the solid lines are from
Eq. (\ref{f_es_expan_bcp}) augmented with higher order terms in $\Delta\eps$ and
$\varphi_{\bf q}$. The higher-order corrections in $\Delta\eps$ are essential since an
interchange of the dielectric constant of the A and B polymers does {\it not} leave the
electrostatic energy invariant.
As the electric field is increased from
zero, the area occupied by the distorted bcc phase shrinks in favor of the
disordered and hexagonal phases. The phase lines terminate at a triple point 
($\chi_{\rm tr}$, $E_{\rm tr}$). The bcc phase completely disappears when
$E>E_{\rm tr}$. The transition from spheres to cylinders 
was also studied for BCPs confined to thin films using SCFT. The
deformation of the spheres from their perfect shape and the 
stability diagrams were calculated for the various system parameters
\cite{matsen_jchemphys2006}.

In the above calculation, the zero-field critical point at
$(N\chi$,$\phi_0)\simeq(10.49,1/2)$ is
left unchanged because of the linear relation $\eps(\phi)$ and because lamellae and
cylinders suffer no electrostatic penalty once oriented along the external field.
This is not true, however, if the constitutive relation
has a nonvanishing quadratic dependence.
To see that this dependence can change the ODT point, consider $\eps''\neq 0$ in
Eq. (\ref{const_relation}), and 
$\varphi=\vphi_L\cos(qx)$ [see Eq. (\ref{op_fourier})], $\phi_0=1/2$
(symmetric lamellae), and ${\bf E}_0$ is in the $y$ or $z$ directions. 
We find that $f_{\rm es}=-(1/2)\bar{\eps}E_0^2-(1/8)\eps ''\vphi_L^2E_0^2$.
The electrostatic energy of the disordered phase is $f_{\rm
es}=-(1/2)\bar{\eps}E_0^2$, and therefore the electrostatic energy favors
oriented lamellae over disorder if $\eps''>0$; on the other hand, when $\eps''<0$
lamellae are expected to melt for large enough field even if they are oriented in the
``right'' direction.

Lastly, we mention that the non-mean-field effect of uniform electric fields on
composition fluctuations in symmetric block copolymers has been calculated as
well. The weak first-order phase transition occurring in short polymers is
shifted closer to the mean-field value, that is, the ODT point is shifted to
higher temperatures \cite{stepanow_tta_mm2007,stepanow_ttalbrecht_pre2009}. As a result,
phase separation is
favored even if $\eps''=0$.
The shift to the ODT temperature was estimated to be about $2.5$ K.
A systematic measurement of the constitutive
relation $\eps(\phi)$ is required in order to compare theory and
experiment and to distinguish which of the terms -- $\Delta\eps$ leading to
mixing, fluctuation effect leading to demixing, or $\eps''$ leading to
either mixing or demixing -- is dominant in order-order phase transitions in
polymer systems.

\section{Liquid Mixtures in Electric Field Gradients} \label{sect_efips}

Up to this point we have been primarily interested in uniform fields. Since even inside an
ideal parallel-plate condenser composition inhomogeneities lead to nonuniform
fields, we defined uniform fields as those present in the system if the
dielectric constant is uniform. In the language of Eq. (\ref{E_expansion}), this means
${\bf E}_0$ is constant. 
Spatially nonuniform fields have important consequences: they break the 
translational and the rotational symmetries, and therefore the symmetry
of the phase diagram around $\phi_c$ is also broken.
For a bilayer of two liquids in perpendicular electric field [Fig.~
\ref{fig_liquid_para_perp}(a)] a uniform field ${\bf E}_0$ deforms the interface
only above a certain value, whereas a nonuniform field ${\bf E}_0$ deforms an
interface no matter how small its strength is.
Mathematically, when ${\bf E}_0$ is nonuniform 
the linear terms in $\vphi$ in
the expansion of $f_{\rm es}$ [Eq. (\ref{f_es_expansion})] survive the
integration. The resulting dielectrophoretic force tends to ``suck'' material 
with high
$\eps$ toward regions with high values of $E^2$. 

As before, we stick for concreteness to a binary mixture of two
simple liquids, denoted 1 and 2, with similar molecular volumes. 
The conclusions we draw
are valid with only small modifications also to liquid-vapor coexistence in
field gradients.
The mixture free energy
as a function of composition, $f_m(\phi)$, is taken to be symmetric with
respect to $|\phi-1/2|$. The mixture has an upper consolute point; 
if $T>T_c$ the energy is convex at all compositions and the
mixture is homogeneous. Below $T_c$, $f_m$ has a double-minimum shape. The transition
(binodal) curve $\phi_t(T)$ is given by $df_m(\phi_t,T)/d\phi=0$.

We start
with neat dielectric liquids, and in Sec. \ref{efips_ions} we generalize to
ion-containing liquids. 

\subsection{Ion-free mixtures}\label{efips_neat}

The mixture is subject to electric field emanating from a collection
of conductors $1$, $2$,~.~.~.~,~$N$, with prescribed potentials $\psi_1$,
$\psi_2$,~.~.~.~,~$\psi_N$, or charged with charges $Q_1$, $Q_2$,~.~.~.~,~$Q_N$ each.
We are interested in finding the potential distribution $\psi({\bf r})$ and composition
profile $\phi({\bf r})$ as a function of the external potentials or charges.  
The total free energy density is given by
\begin{eqnarray}
f=f_m(\phi)-\frac12\eps(\phi)\left(\nabla\psi\right)^2-\mu\phi~.
\end{eqnarray}
The equilibrium profiles are governed by the following Euler-Lagrange equations:
\begin{eqnarray}\label{efips_gov_eqns}
\frac{\delta f}{\delta\phi}&=&\frac{\delta f_m}{\delta\phi}-\frac12\frac{d\eps}
{d\phi}\left(\nabla\psi\right)^2-\mu=0~,\\
\frac{\delta f}{\delta\psi}&=&\nabla\left(\eps(\phi)\nabla\psi\right)=0~.\nn
\end{eqnarray}
In these equations, $\mu$ is the chemical potential needed to conserve the
average mixture
composition $\phi_0$, 
\begin{eqnarray}\label{efips_conserv}
\mathcal{V}^{-1}\int\phi({\bf r}){\rm d}^3r&=&\phi_0~,
\end{eqnarray}
where ${\mathcal V}$ is the volume or is the reservoir chemical potential in case of a
grand-canonical ensemble.
The second variational equation is simply the Laplace equation. The set of 
differential equations 
[Eq. (\ref{efips_gov_eqns})], are coupled and nonlinear and quite difficult to solve. 
Above $T_c$, $f_m(\phi)$ is convex. We therefore expect that under the spatially
nonuniform forcing of the external field the composition $\phi({\bf r})$ will be a
smoothly varying
function of space. Variations in $\phi$ will be proportional to the
variations in field, and if $E_0({\bf r})$ is continuous, so is $\phi({\bf r})$. 

\begin{figure}[h!]
\includegraphics[scale=0.45,bb=48 630 525 760,clip]{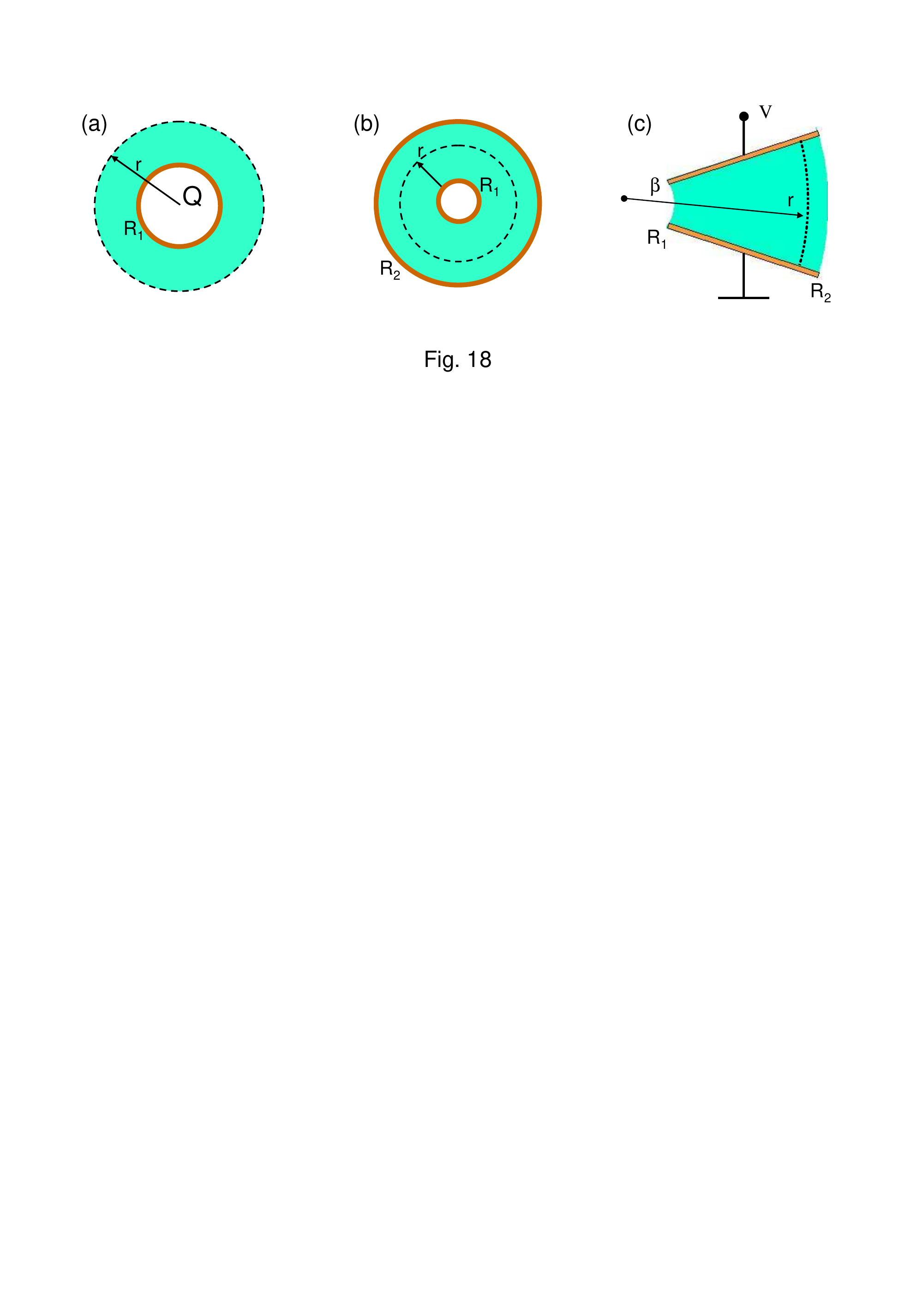}
\caption{Three model systems where field gradients lead to demixing. (a) A single charged
colloid of charge $Q$ and radius $R_1$. (b) A charged wire with charge density $\lambda$
per unit length and radius $R_1$ or two concentric cylinders with radii $R_1$ and $R_2$.
(c) A wedge comprised of two flat electrodes with an opening angle $\beta$ and potential
difference $V$. $R_1$ and $R_2$ are the minimal and maximal values of the distance $r$
from the imaginary meeting point.
}
\label{fig_three_geometries}
\end{figure}

Below $T_c$, this is
not true, and one can expect a different behavior as is shown below. 
Consider three
``canonical'' examples illustrated in Fig.~\ref{fig_three_geometries}. 
The first one is a charged isolated spherical colloid of radius
$R_1$ and total charge $Q$ coupled to a reservoir at infinity. 
Azimuthally symmetric profiles depend on $r$ only, where $r$ is
the distance from the colloid's center. Thus 
$\phi=\phi(r)$, $\eps=\eps(r)$, and the field is given by
${\bf E}(r)=Q/\left(4\pi\eps[\phi(r)]r^2\right)\hat{\bf r}$.
The second geometry is a charged wire of radius
$R_1$ and charge $\lambda$ per
unit length, coupled to a reservoir at $r\to\infty$. Alternatively, we may
consider a
closed condenser made up of two concentric cylinders with fixed average mixture
composition. 
In both cases the electric field is ${\bf
E}(r)=\lambda/\left(2\pi\eps[\phi(r)]r\right)\hat{\bf r}$.
A third
system is the so-called ``wedge,'' made up from two flat electrodes with an
opening angle
$\beta$ between them, with potential difference $V$. In the wedge, ${\bf E}$ is given
by ${\bf E}=V/(\beta r)\hat{{\bf \theta}}$,
irrespective of the profiles $\phi(r)$ or $\eps(r)$. The dielectrophoretic
force existing in the wedge condenser may lead to discontinuous
meniscus location in immiscible liquids \cite{jones_jap1974}, or 
to exact cancellation of gravity effects in mixtures near their critical
point \cite{tsori_pre2005}.

Nonuniform electric fields will not demix liquids 
such that liquid interfaces are created parallel to the field.
To see this, consider the two concentric cylinders with potential
difference $V$. 
In the homogeneous state,
$\eps=\bar{\eps}=\phi_0\eps_1+(1-\phi_0)\eps_2$ is uniform.
The demixed state has two uniform liquids of permittivities $\eps_1$ and
$\eps_2$ occupying two portions of relative volumes $\phi_0$
and $1-\phi_0$, respectively, and the interfaces 
are parallel to ${\bf E}$. 
The homogeneous and demixed states have
exactly the same electric field and same
value of $F_{\rm es}$, but $F_m$ is higher for the demixed state. 
If the inner cylinder has a charge
$\lambda$ per unit length {\it immobilized}
to it, then the difference in $F_{\rm es}$ 
between the demixed and the homogeneous states is 
$\lambda^2/(4\pi)\ln(R_2/R_1)[\phi_0/\eps_1+(1-\phi_0)/\eps_2-1/\bar{\eps}]$, and this 
difference is always positive.

The creation of interfaces perpendicular to the external field is in
contrast to our intuition about uniform electric fields: despite the effects of 
dielectric interfaces, the dielectrophoretic force is the dominant mechanism.
In addition, when a nonuniform field acts parallel to a liquid interface, it destabilizes
it
\cite{jones_jap1972,jones_jap1975}.

Additional comment can be made on the possible mobility of charges inside the
conductors: 
consider a single flat conductor
at $x=0$, charged with average charge $\bar{\sigma}$ per unit area (uniform fields). 
The mixture is at the half space
$x>0$ and has average dielectric constant $\bar{\eps}$. 
Is it beneficial for the system to separate into two regions, one
at $y>0$, with surface charge $\bar{\sigma}+\delta\sigma$ (at $x=0$) and
dielectric constant $\bar{\eps}+\delta
\eps$ (at $x>0$), and one at $y<0$, with surface charge
$\bar{\sigma}-\delta\sigma$ ($y=0$) and dielectric
constant $\bar{\eps}-\delta\eps$ ($x>0$)? Even without accounting for the
reduced entropy of charges,
this will not occur since the difference in $F_{\rm es}$ between demixed and 
homogeneous states is proportional to
$(\delta\sigma/\bar{\sigma}-\delta\eps/\bar{\eps})^2$
and is always positive.
Patches can be created on 
the conductor's surface, though, if the charges have effective repulsion due to van der
Waals
forces \cite{safran_pincus2007}.

Let us return to demixing in nonuniform fields. 
The nice thing about the three examples in Fig.~\ref{fig_three_geometries},
besides the geometrical simplicity,
is the
{\it decoupling} between the two equations [Eqs. (\ref{efips_gov_eqns})]. 
Essentially, we
have already solved for Laplace's equation and only need to substitute the value of
$E$ in the variation with respect to $\phi$. We are left with a single nonlinear
equation for $\phi(r)$, which has spatial dependence. Let us ignore for the moment the
term proportional to $(\nabla\phi)^2$ in $f_m$. This can be justified because we are
dealing with a volume term in $E^2(r)$. For a large system, 
the surface tension term acts only at sharp interfaces and can be neglected as a first
approximation. For smaller systems, surface tension smooths out the composition profile
$\phi(r)$ and the neglect cannot be justified.

In order to be concrete, consider the wedge geometry coupled to a reservoir at 
composition
$\phi_0$ ($R_2\to\infty$) and assume that $\eps$ varies linearly 
with $\phi$ [$\eps''=0$
in Eq.(\ref{const_relation})]. We find the equation for $\vphi$ is
\begin{eqnarray}\label{wedge_gov_eqn}
f_m'(\vphi)=\frac12\Delta\eps\left(\frac{V}{\beta r}\right)^2+\mu
\end{eqnarray} 
(recall that $\vphi=\phi-\phi_c$ is the deviation from critical composition).
In the absence of field, the composition is $\phi_0$, corresponding to a homogeneous
phase above the binodal. With electric field, Eq. (\ref{wedge_gov_eqn}) gives an
algebraic relation between $\vphi$ and $r$. Above $T_c$, the left-hand side is a
monotonously increasing function behaving qualitatively as $\vphi+\vphi^3$.
Therefore, $\vphi$ increase monotonically 
as $r$ decreases. This is the regular dielectrophoretic effect. 

However, below $T_c$, 
the left-hand side has a sigmoidal shape; it behaves qualitatively as
$-\vphi+\vphi^3$.
One may plot $f_m'(\vphi)$ against $\vphi$ at a fixed temperature.  
The graphical solution to Eq. (\ref{wedge_gov_eqn}) is then given as an
intersection between this curve and a horizontal line independent of $\vphi$.
At large values of $r$ (small electric field), the intersection occurs at
negative values
of $\vphi$. As $r$ decreases, $\vphi(r)$ increases. If the maximum field
in the system $V/(\beta R_1)$ is small, then the profile is still monotonic. If,
however, $V$ is large enough, the solution to $\vphi$ to Eq.
(\ref{wedge_gov_eqn}) occurs at negative
values for large $r$'s and positive values for small $r$'s (large field).
Necessarily, there is a location $r$ where $\vphi$ has a ``jump.''
This is the field-induced phase transition: at a
certain value of the voltage, the profile becomes {\it discontinuous}
\cite{ttl_nature2004,marcus_jpsj2009}. 
After demixing
occurs, the wedge exhibits two regions with different compositions separated by a sharp
interface. 
The two compositions of the coexisting phases are themselves nonuniform.

The interface separating high and low values of $\phi$ is at $r=R$
($R_1<R<R_2$).
The critical 
potential $V^*$ is the minimal potential for demixing, and it occurs when $R$ attains its 
smallest value, i.e., $R=R_1$.
As the potential increases,
the interface moves to larger radii, and, in addition, the composition difference between
the coexisting domains increases. For a closed wedge with average composition $\phi_0$,
one 
readily finds that $R(V)$ is bounded, and the maximum value of $R$ is given by 
$R_{\rm max}^2=(1-\phi_0)R_1^2+\phi_0 R_2^2$. 
If the wedge is coupled to an 
external reservoir and $R_2=\infty$, $R$ grows indefinitely as a 
function of $V$.

\begin{figure}[h!]
\includegraphics[scale=0.5,bb=38 575 535 795,clip]{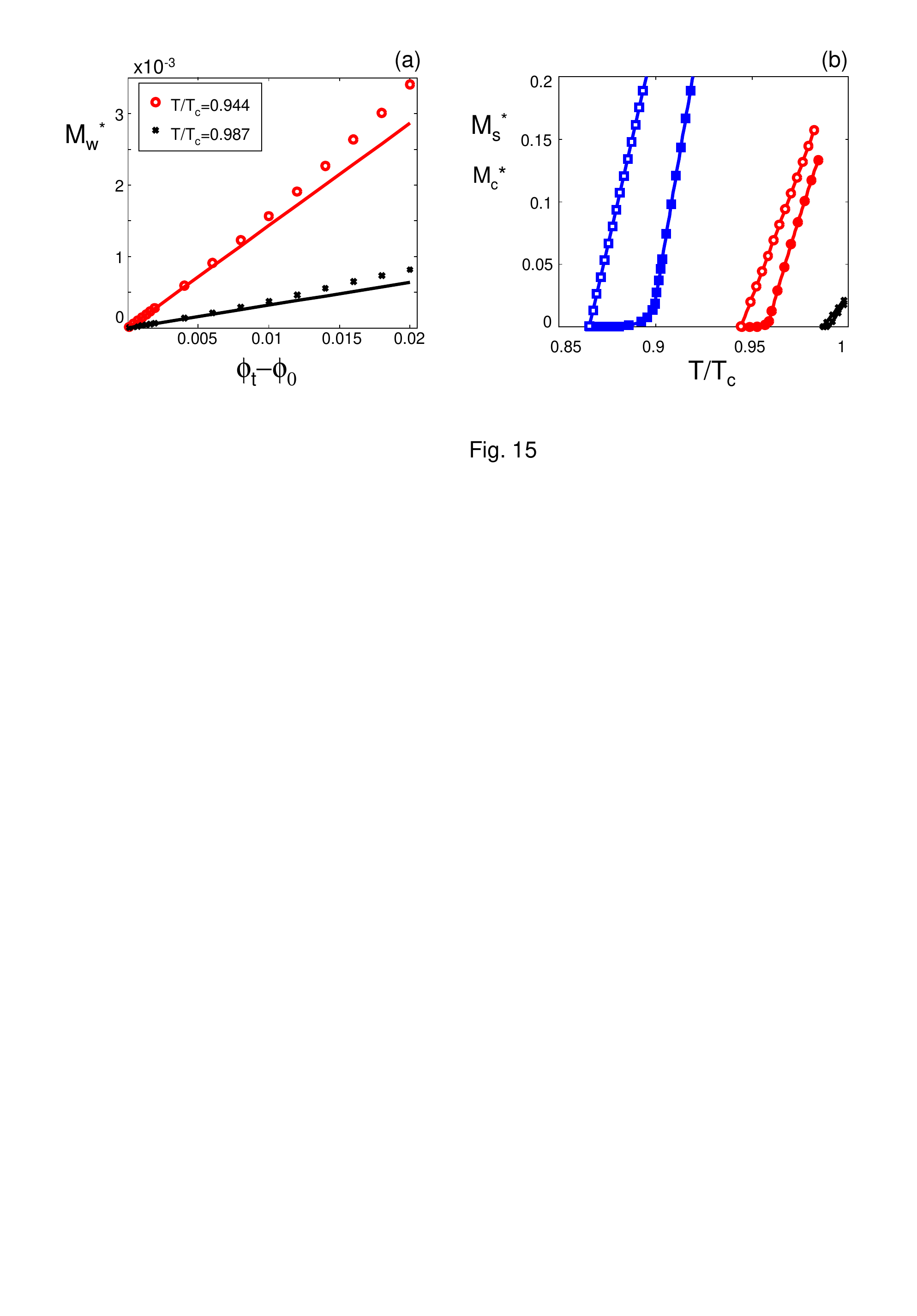}
\caption{Critical demixing fields. 
(a) Scaled critical voltage for demixing $M_w^*$ for a closed wedge
as a function
of distance from the transition (binodal) composition $\phi_t$. Symbols are numerical values 
and line is approximate analytical expression [Eq. (\ref{Mw_approx})].
(b) Scaled critical charge of spherical and cylindrical
colloids as a function of temperature. Open symbols are $M_c^*$ for a closed cylindrical
system of compositions $\phi_0=0.2$ (squares), $\phi_0=0.3$ (circles), and $\phi_0=0.4$
(crosses). Filled symbols are $M_s^*$ for an isolated spherical colloid coupled to a 
reservoir with the same compositions.
Reprinted with permission from \textcite{marcus_jcp2008}. Copyright 2008,
American Institute of Physics.
}
\label{fig_crit_Ms}
\end{figure}

The dimensionless potential $M_w$ can be written as 
$M_w\equiv V^2Nv_0\eps_0/4\beta k_BT_cR_1^2$. Here we generalize to mixtures
of polymers, 
$N$ is the polymerization index and $Nv_0$ is the total chain volume.
Close enough to the transition composition $\phi_t$, 
the critical value of $M_w$, $M_w^*$, is bounded by the following expression:
\begin{eqnarray}\label{Mw_approx}
M_w^*=\frac{\phi_t-\phi_0}{4\Delta\tilde{\eps}}\frac{T}{T_c}
\frac{d^2\tilde{f}_m(\phi_t)}{d\phi^2}g(R_2/R_1)~,
\end{eqnarray}
where we have used $\Delta\tilde{\eps}=\Delta\eps/\eps_0$, and $\tilde{f}_m=
Nv_0f_m/k_BT$. $g(x)=2(x^2-1)/(x^2-1-2\ln x)$ is a geometry-specific
dimensionless 
function of the ratio $x\equiv R_2/R_1$.
Figure~\ref{fig_crit_Ms}(a) shows $M_w^*$ as a function of difference between the
transition
composition $\phi_t$ and $\phi_0$. Solid lines are Eq. (\ref{Mw_approx}), while symbols
are 
taken from a more accurate numerical calculation.
Clearly, $M_w$ increases almost linearly with $\phi_t-\phi_0$. The slope
decreases at $T$ approaches $T_c$.
Note that the similarity between the wedge
capacitor and a rapidly rotating centrifuge allows us to predict phase-separation
transitions for mixtures in centrifuges if the rotation frequency is
higher than a certain critical frequency \cite{tsori_crphysique2007}.

The behavior in the cylindrical (2D) and spherical (3D) cases is similar to the one
described
above for the wedge. There are 
expressions for the critical values of the dimensionless charges
for the cylindrical and spherical symmetries: $M_c\equiv \lambda^2
Nv_0/16\pi k_BT_cR_1^2\eps_0$ and $M_s\equiv Q^2Nv_0/64\pi
k_BT_cR_1^4\eps_0$, respectively.
Figure~\ref{fig_crit_Ms}(b) shows the critical values for demixing, $M_c^*$ and $M_s^*$,
as a function of temperature for three
different compositions. Open symbols are calculated values of $M_c^*$ for two concentric
cylinders, whereas filled symbols are $M_s^*$ for an isolated spherical colloid.
For a given composition, curves only start above the transition temperature ($T/T_c\simeq
0.865$, $0.944$, and $0.986$ for $\phi_0=0.2$, $0.3$, and $0.4$, respectively).
Note that in general, $M_s^*$ and $M_c^*$ are much larger than $M_w^*$ -- 
this is again because
of the dielectric interfaces: in the spherical and cylindrical symmetries, ${\bf E}$ is
parallel
to $\nabla\phi$, and therefore the dielectrophoretic force $\propto\Delta\eps$ has to be
large
enough in order to overcome the penalty associated with dielectric interfaces, 
$\propto(\Delta\eps)^2$. In the wedge ${\bf E}\perp\nabla\phi$, and $M_w^*$ is
correspondingly
smaller.

How much should a $1$ $\mu$m colloid be charged to create a demixing layer around it? We take
the
molecular volume $v_0$ to be $3\times 10^{-27}$ m$^3$, $N=1$ (simple liquids), and 
$T_c=350$ K. We find that 
$M_s^*=0.001$ leads to $Q\sim 10^4e$. This may seem like a large number; however, if the 
mixture contains even a low molecular weight polymer, then $N=500$ gives $Q\sim 470e$.
The enrichment layer created around the colloid may be quite important for the thermodynamic
behavior of suspensions because field-induced demixing leads to capillary
attraction, promoting colloidal aggregation.

Finally, for a given value of $\phi_0$, there is an approximate expression for the range of 
temperatures $\Delta T$ above the transition temperature
$T_t(\phi_0)$ for which a given voltage is able to demix,
\begin{eqnarray}\label{DT_efips}
\Delta T=\frac{Nv_0}{2k_B}\left|\frac{\Delta\eps}{\phi_c-\phi_0}\right|\left(\frac{V}
{\beta R_1}\right)^2~.
\end{eqnarray}

This formula, applicable to the wedge, is valid close enough to the binodal. It
should be 
contrasted with the Landau expression [Eq. (\ref{DT_landau_mechanism})]. When the
factor $N/2$ is put aside 
the two formulas have similar form: $\eps'' E_0^2$ in Eq.
(\ref{DT_landau_mechanism}) is 
replaced by $\Delta\eps(V/\beta R_1)^2$, and both expressions have similar
magnitudes since $\Delta\eps/\eps''\sim 1$. However, the denominator of Eq.
(\ref{DT_efips}) has the difference
between $\phi_0$ and the critical composition. This factor means that $\Delta T$
here is at least twice as large as in the Landau case and typically is much
larger: $|\phi_c-\phi_0|^{-1} \sim 20$ or even more.

The phenomenon described here has many similarities to regular
prewetting. Important differences are (i) ${\bf E}$ is determined from 
Laplace's equation and a similar equation does not exist for the case of
long-range van der Waals forces between liquids and solid surfaces. (ii) In
addition, the 
electric field needs not attain its highest value at the surface.
Thus, ``electroprewetting'' is a special case of field-induced demixing.
Consider an analogous system of a body of nonuniform heat conductivity
$\kappa({\bf r})$ enclosed by a surface $\Sigma$ with prescribed temperature
$T_\Sigma$. Temperature is the analog of $\psi$ and $\kappa$ is the
analogue of $\eps$.
A question follows: is it possible to have a
distribution $\kappa({\bf r})$ such that the steady-state temperature profile
$T({\bf r})$ achieves a maximum or a minimum at the bulk rather than at 
$\Sigma$? A
positive answer would mean that a confined mixture could spontaneously
demix due to heat flow even without Soret or other effects.
The answer to the question is negative. Since $T$ is governed by
$\nabla(\kappa \nabla T)=0$, we find $\nabla\kappa\cdot\nabla
T+\kappa\nabla^2T=0$, and extrema do not satisfy this equation because $\nabla
T=0$ and $\nabla^2T\neq 0$.
However, the point here is that electric fields are different: the fact that
$\psi$ is largest on the electrodes does not prevent $E^2$ from being the
largest elsewhere.

Up to this point, the liquids were considered as pure dielectrics. In the
following section the phase-separation phenomenon is described in the presence
of ions.

\subsection{Phase separation in ion-containing liquids}\label{efips_ions}

The field-induced phase transition described above originates from electric-field
gradients. In nonpolar 
liquids, such gradients occur when the electrodes are curved. In one
dimension, e.g., in parallel-plate condenser, field gradients are forbidden
by the equation $\nabla\cdot {\bf E}=0$. For curved 
electrodes, the geometry dictates the length scale over which the
electric field falls off.
On the other hand, mixtures of polar liquids (e.g., aqueous solutions)
contain some amount of charge carriers. 
In such mixtures, the physics is rich and quite different from the simple
dielectric case. The most important feature is due to {\it screening}, brought
about by dissociated ions accumulating at the electrodes.
Screening means that the electric field is substantial only close to the
electrodes, within the screening distance $\lambda$. Field gradients originate
from both geometry and screening, and the phase transition 
depends on two lengths. The ionic screening thus adds to the
dielectrophoretic
force which separates the liquid components from each other. Since screening is
omnipresent, phase separation may occur even near parallel and flat charged
surfaces, i.e., in one dimension.

But ions have another effect besides increasing the dielectrophoretic force.
Ions have in general different solubilities in the different liquids. 
As an ion drifts toward the electrode, it might ``drag'' with it the preferred
liquid component \cite{onuki_jcp2004}. 
Usually the two ionic species are preferential in the same
liquid, but it is also possible that they are preferentially soluble in different
liquids. The solubility introduces a force of electrophoretic origin,
proportional to the ions' charge.
In binary mixtures, the parameter $\Delta u^+$ measures how much a
positive ion prefers to be in the environment of liquid 2 over that of liquid
1. To lowest order, the interaction per unit volume of positive ions and the
mixture is given by $-\Delta u^+n^+({\bf r})\phi({\bf r})$, where $n^+$ is the
positive ion number density.
A similar expression exists for the negative ions.
The system free energy
density including the interaction of ions and field is given on the mean-field
level by \cite{onuki_pre2006,onuki_jpcb2009}
\begin{eqnarray}
f&=&f_m(\phi)-\frac12\eps(\phi)\left(\nabla\psi\right)^2
+\left(n^+-n^-\right)e\psi\\
&+&k_BT\left[n^+\ln\left(v_0n^+\right)+
n^-\ln\left(v_0n^-\right)\right]-\lambda^+n^+\nn\\
&-&\lambda^-n^--\mu \phi-\left(\Delta 
u^+n^++\Delta 
u^-n^-\right)\phi+{\rm const.}\nn
\end{eqnarray} 
In the above, $\lambda^\pm$ and $\mu$ are the Lagrange multipliers 
(chemical potentials) of the positive and negative ions and liquid 
concentration, respectively. 

The free energy needs to be extremized with respect to the fields $\phi$,
$\psi$, and $n^\pm$ in keeping with a fixed ion concentration
${\mathcal V}^{-1}\int n^\pm({\bf r}){\rm d}^3r=n_0$. The Poisson-Boltzmann equation is 
obtained from the variation in $f$ with respect to the potential and is 
coupled to $\delta f/\delta\phi$. This
formalism is similar to other formulations, but the crucial difference
is the bistability of the mixture energy $f_m$ \cite{onuki_jcp2004,andelman_jpcb2009}.

Solution to the problem is difficult to obtain because of the essential
nonlinearity: a linear Poisson-Boltzmann equation does not suffice since
imposing $e\psi\ll k_BT$ means small potential and thus forbids the large
electric fields necessary to drive the phase-separation transition.
The transition temperature $T_t(\phi_0)$, defined by
$df_m(\phi_0,T_t)/d\phi=0$, 
is shifted by an amount $\Delta T$ due to the external potential. 
An approximate formula for $\Delta T$
near a single planar electrode at $x=0$ can be obtained by the assumption that
field gradients are dominated by ionic screening rather than by $\eps$
variations. It reads \cite{tsori_pnas2007}
\begin{eqnarray}\label{DT_efips_ions}
\frac{\Delta T}{T_c}\simeq\left(\frac{|\Delta\eps|}{\eps_c}+ \frac{\Delta
u}{k_BT_c}\right)\frac{Nn_0v_0}{|\phi_c-\phi_0|}
\exp\left(\frac{eV}{k_BT_c}\right)~, 
\end{eqnarray}
where $\Delta u=\Delta u^+=-\Delta u^-$ and $V$ is the electrode potential.
This expression holds as long as $T_t+\Delta T$ is smaller than $T_c$; at all
temperatures $T>T_c$, the composition profile $\phi(x)$ varies smoothly
with no abrupt jump. The shift in $T_t$ should be compared to Eq.
(\ref{DT_efips}) for a mixture without ions and to Eq.
(\ref{DT_landau_mechanism}) for a mixture in uniform field.
The dielectrophoretic and electrophoretic driving forces
appear to have similar significance since both $\Delta\eps/\eps_c$ and 
$\Delta u/k_BT$ are $\sim O(1)$. The numerator has a small factor $Nn_0v_0$; 
in pure water and if we take $Nv_0=3\times 10^{-27}$ m$^3$, we get
$Nn_0v_0\sim 10^{-7}$. Small as this factor is, and even ignoring the
denominator of $|\phi_c-\phi_0|$, the exponential factor $\exp(eV/k_BT)$ 
is the dominant term in $\Delta T$. It is usually huge, recalling that
at room temperature $1$ eV equals $40k_BT$.

One can therefore expect that 
in almost all circumstances the electroprewetting discussed above
should occur around a charged colloid in aqueous solutions. 
The width of wetting layer is small but may nonetheless influence the effective
interaction potential between two colloids or between a colloid and a
nearby surface \cite{dietrich_pre2000} and the
thermodynamic behavior of suspensions of charged particles
\cite{beysens_prl1985}. Electric field gradients may also play a role in such
phenomena as the ``critical Casimir effect,'' recently reported experimentally
by \textcite{bechinger_nature2008}.

The above wetting transition from a homogeneous to a demixed solution has
several aspects. First, if the two or more constituent pure components have
sufficiently differing indices of refraction, the optical properties of the
solution will be drastically different in the two states. The
demixing of an initially transparent solution leads to the creation of optical
interfaces, and these can be used to scatter, reflect, and refract light.
Particularly in a microfluidic environment, the phase separation could be
utilized to create waveguides which are liquid and switchable
\cite{psaltis_nature2006}. The mechanical properties of the mixture are also
different whether it is homogeneous or not. If the two pure components have 
different viscosities, as is the case of the experiments \cite{ttl_nature2004},
then after demixing a thin lubricating layer is created at the charged
surfaces, and this thin layer completely alters the flow profile between the
surfaces. Finally, the field-induced phase separation can be used to control
chemical reactions of, say, two molecules A and B reacting in a mixture of
liquids 1 and 2. The dependence of concentration on electric field can
accelerate the reaction or conversely stop it. The spatial dependence can be
used to confine the reaction to small volumes or to two-dimensional interfaces
\cite{tsori_pnas2007}.

\subsection{Phase separation in surfactant mixtures}\label{efips_surfactants}

Phase transitions in nonuniform electric fields are not restricted to
demixing in liquid mixtures or to a liquid-vapor coexistence. In a set
of beautiful experiments, \textcite{kycl_science1994} demonstrated that a flat
monolayer of a binary lipid mixture on the water-air interface undergoes a
transition from a homogeneous to a phase-separated state when an insulated wire
passing perpendicular to the monolayer is charged. The spatially varying
electric field couples to the different dipole moments of the constituent
molecules and attracts the molecules with large dipole toward the wire.

These experiments have an important difference from the demixing described
before: the electric field couples to the fixed dipole of the lipids. Thus, the
effect scales
linearly with $E$ rather than quadratically. Consequently, when the wire
potential is $V$, lipid 1 is attracted to the wire, whereas the
potential reversal $V\to -V$ reverses the effect and accumulation of lipid 2
is observed at the wire. Note that the dipole orientation is restricted to be
in one direction; if dipoles could flip upside down, a reversal of the field
$E\to -E$ would leave the system invariant and the effect would again scale as
$E^2$. Similarly to the demixing in simple liquids, below the
critical pressure (which replaces $T$ in the phase diagram) only smooth
gradients in $\phi$ appear, whereas below the critical pressure true demixing
occurs, as evidenced by the sharp interface separating coexisting domains.

\section{Conclusions}	

In this Colloquium we considered 
liquids and polymers in external electric fields. 
The key quantities for the physical behavior are the linear and quadratic
coefficients in the
expansion of $\eps(\phi)$, namely, $\Delta\eps$, and $\eps''$ in Eq.
(\ref{const_relation}), and the conductivity
$\sigma$. We discussed the significance of these quantities in several phenomena
occurring when the field is either spatially uniform or nonuniform.
We have
emphasized that in a uniform field ${\bf E}_0$, there 
is a preference of electrostatic origin to have
dielectric interfaces parallel to ${\bf E}_0$. This preference,
depending on $\Delta\eps$, gives rise to
the perpendicular field instability observed for liquid films discussed in
Sec. \ref{interf_instab}. The wavelength of instability $\lambda^*$ is the
fastest-growing mode of a dynamical evolution.
What is the smallest feature size possible in
these experiments? The answer is not clear, and an effort is undertaken by
several groups to
minimize $\lambda^*$ by optimizing the fluid's surface tension, film thickness,
and other parameters.
Dielectric interfaces are also responsible for block-copolymer
orientation and order-order phase transitions in electric fields. Current
effort focuses on finding ways to eliminate in-plane defects and enlarging the
grain size.
Destabilization of copolymer interfaces and polymer brushes may
also occur, but here the wavelength is dictated by equilibrium considerations.
In addition, a uniform electric field deforms the shape of suspended drops and
of drops on solid substrates.

We have outlined some of the effects related to dissociated ions, commonly
present in most simple and polymeric liquids. Finite conductivity introduces
additional stress and can lead to oblate shapes of an isolated drop.
Electrowetting is a result of charge accumulation at the droplet-substrate
interface. The
reduction in the droplet-substrate interfacial tension leads to an apparent
change in 
the contact angle, and this has been widely used in applications.
Ions also play an important  role in copolymer orientation and
phase transitions. Complexed ions mainly contribute to enhancement of
$\Delta\eps$, whereas mobile ones may give rise to wettinglike effects due to
their preferential solubilities. An experimental effort is underway, aiming to
understand when complexed ions are more dominant than mobile ones and vice
versa. The study of electric-field-induced wetting of complex phases, such as
block copolymers, is still at early stages and deserves experimental and 
theoretical treatments.
Ions introduce stresses when the fields are time varying, and the resulting
torques tend to orient the copolymer domains. The dynamics of these
out-of-equilibrium systems is an open problem that also requires further
investigation.

Uniform electric fields can also lead to phase transitions in liquid mixtures
by the Landau mechanism. These transitions are insensitive to $\Delta\eps$ 
but rely on nonvanishing $\eps''$. Depending on the sign of $\eps''$, the whole
binodal curve can be shifted to higher or lower temperatures, but this shift is
rather small, $\sim 0.03$ K.

Spatially nonuniform fields deform ordered block-copolymer phases due to a
nonvanishing $\Delta\eps$. The copolymer structure, a compromise between 
elastic bending and compression energies and the electrostatic energy, has not
received enough attention so far. It is hoped that research in this direction
can show us how to eliminate defects and perhaps even how to tailor
lamellae or other phases in a prescribed way on a substrate.
In the much simpler systems of liquid mixtures, bending is irrelevant, and
nonuniform fields have a strong effect on the phase behavior. The 
direct coupling between field and composition leads to a phase transition which
depends on $\Delta\eps$ and cannot be simply described by a renormalization
of $T_c$. The change in transition temperature is much larger than in the Landau
mechanism. Wetting layers are created around curved charged objects but can
also appear far from a solid interface. 
The presence of dissociated ions greatly
enhances the transition and allows it to happen at flat objects.
This transition can have numerous applications in microfluidics, 
in micro and nanorheology and lubrication, in
chemical reactions, etc.
The current study is only at its infancy and there is still plenty
to learn: for example, what is the size and velocity of liquid domains
appearing upon switching the field on, why interfacial instabilities are
observed after demixing takes place, 
what is the role of critical
fluctuations and the non-mean-field behavior, how does heating affects the
transition, etc.

The phenomena  outlined above are fundamentally interesting and also have
numerous applications. We are certainly going to learn more about this active
field of research in the years to come.

\section*{Acknowledgments}

I would like to thank L. Leibler and F. Tournilhac for insightful discussions
and for help in developing the ideas presented in this work and D. Andelman and
M. Schick for critical comments and fruitful discussions. Communications
from A. B\"{o}ker, C. Bechinger, S. Dietrich, S. Herminghaus,  
M. M\"{u}ller, F. Mugele, A. Onuki, 
W. B. Russel, T. P. Russell, U. Steiner,  S. Stepanow, and T. Thurn-Albrecht
are gratefully acknowledged.
I acknowledge support from the Israel Science Foundation (ISF) Grant No. 284/05,
the German-Israeli Foundation (GIF) Grant No. 2144-1636.10/2006, and 
the COST European program P21 ``The Physics of Drops''.

\bibliographystyle{apsrmp}
\bibliography{rmp_colloquium_condmat}

\end{document}